\documentclass[twocolumn]{emulateapj}

\shorttitle{Glitch analysis of stars in {\it Kepler} seismic LEGACY sample}
\shortauthors{Verma et al.}

\begin{document}

\title{Seismic measurement of the locations of the base of convection zone and helium ionization zone for stars in the 
{\it Kepler} seismic LEGACY sample}

\author{Kuldeep Verma\altaffilmark{1,2,3}, Keyuri Raodeo\altaffilmark{4}, H.~M.~Antia\altaffilmark{2},
Anwesh Mazumdar\altaffilmark{4}, Sarbani Basu\altaffilmark{5}, Mikkel N.~Lund\altaffilmark{6,1}, 
V\'{i}ctor Silva Aguirre\altaffilmark{1}}

\altaffiltext{1}{Stellar Astrophysics Centre, Department of Physics and Astronomy, Aarhus University, Ny Munkegade 120, DK-8000 
Aarhus C, Denmark}
\altaffiltext{2}{Tata Institute of Fundamental Research, Homi Bhabha Road, Mumbai 400005, India}
\altaffiltext{3}{Centre for Excellence in Basic Sciences, University of Mumbai, Kalina, Mumbai 400098, India}
\altaffiltext{4}{Homi Bhabha Centre for Science Education, TIFR, V.~N. Purav Marg, Mankhurd, Mumbai 400088, India}
\altaffiltext{5}{Astronomy Department, Yale University, P.~O. Box 208101, New Haven, CT 065208101, USA}
\altaffiltext{6}{School of Physics and Astronomy, University of Birmingham, Edgbaston, Birmingham, B15 2TT, UK}

\email{kuldeep@phys.au.dk}

\begin{abstract}
Acoustic glitches are regions inside a star where the sound speed or its derivatives change abruptly. These leave a small
characteristic oscillatory signature in the stellar oscillation frequencies. With the precision achieved by {\it Kepler} seismic 
data, it is now possible to extract these small amplitude oscillatory signatures, and infer the locations of the 
glitches. We perform glitch analysis for all the 66 stars in the {\it Kepler} seismic LEGACY sample to derive the locations of the 
base of the envelope convection zone and the helium ionization zone. The signature from helium ionization zone is found to 
be robust for all stars in the sample, whereas the convection zone signature is found to be weak and problematic, particularly
for relatively massive stars with large errorbars on the oscillation frequencies. We demonstrate that the helium glitch 
signature can be used to constrain the properties of the helium ionization layers and the helium abundance.
\end{abstract}


\section{Introduction}
\label{sec:intro}
An accurate understanding of stellar structure and evolution is of paramount importance to astrophysics, and physics in general. 
Indeed, the properties of stars are used to infer the nature of the associated exoplanets and to learn the history of the Milky
Way. Seismic data from {\it CoRoT} \citep{bagl06,bagl09} and {\it Kepler} \citep{boru09,koch10} space missions have 
revolutionized our understanding of stellar structure. The precise set of observed oscillation frequencies are being used to 
test the various hypotheses of stellar physics. 

It had been proposed that the signatures of the acoustic glitches in the oscillation frequencies of distant stars could be 
used to determine the depths of the base of the envelope convection zone and helium ionization zone \citep{mont00,mazu01,goug02,
roxb03}. The amplitude of the glitch signature is a few orders of magnitude smaller than the background smooth component, and 
a set of precise oscillation frequencies are required in a sufficiently large frequency range to use this technique. With the 
availability of the high quality seismic data from {\it CoRoT} and {\it Kepler} space missions, it has become possible to 
apply this technique to distant stars. \citet{migl10} used the modulation of the frequency separation observed in {\it CoRoT} 
data to determine the location of the helium ionization zone in a red giant. Similarly, \citet{mazu12} used {\it CoRoT} data for 
HD49933 to determine the acoustic depths of the helium ionization zone and base of the envelope convection zone
\citep[see also,][]{mazu11,roxb11}. \citet{mazu14} performed a more extensive study using data from about a year of observations 
by {\it Kepler} for 19 stars to determine the acoustic depths of both glitches. In this work, we extend the study to 66 stars 
using {\it Kepler} data, covering up to 3.5 years of observations. Apart from location, the oscillatory signal from the 
helium ionization zone was shown to be sensitive to the envelope helium abundance \citep{basu04,mont05,houd07}. \citet{verm14a} 
have used oscillation frequencies from {\it Kepler} to determine the envelope helium abundance of a binary system, 16 Cyg A \& B.

The independent measurement of the locations of the base of the convection zone and helium ionization zone can be used to 
constrain the stellar properties better. The fundamental stellar parameters are not only useful in the context of stellar 
evolution but also in the studies of 
exoplanets \citep[see, e.g.,][]{nutz11,gill13,liu14,silv15}, stellar populations \citep{chap11}, and galactic archeology 
\citep[see, e.g.,][]{migl13,casa14,casa16}. The stellar ages are particularly important, and cannot be determined directly.
The standard technique to determine stellar age compares the observed surface properties of the star with the corresponding 
quantities from the stellar evolution models. This approach is effective in determining the ages of stellar clusters, provided 
the observed sample includes stars at different evolutionary stages, including those beyond the main-sequence. But for the 
isolated field stars, we need additional seismic constraints to determine the ages reliably. For instance, the observed 
oscillation frequencies or their appropriate combinations along with spectroscopic data are used to obtain the best-fit stellar 
model \citep[see, e.g.,][]{math12,chap14,metc14,silv15}. Recently, \citet{silv13} have obtained stellar ages to about 10\% accuracy 
using seismic data in addition to spectroscopic observations. The process of determining the best-fit stellar model typically 
involves minimization of a cost function, which is a highly nonlinear function and may have multiple minima 
\citep[see, e.g.,][]{aert10}, and in some cases two or more minima may be close in terms of the quality of the fits. The 
additional information from the acoustic glitches can be used to resolve these near degeneracies.  

The locations of the acoustic glitches can also be used to constrain the input physics of the stellar evolution models. This
was demonstrated by \citet{mazu05} using synthetic data for a {\it CoRoT} target star. The 
evolutionary models require the heavy element abundance, $Z$, of the star, which is generally derived from the observed 
$[{\rm Fe}/{\rm H}]$ assuming the relative abundances to be similar to the solar abundances. The recent revision of the solar 
heavy element abundances using 3D hydrodynamic model for the solar atmosphere \citep{aspl09} are known to be inconsistent with 
the helioseismic constraints \citep[and references therein]{basu08,goug13}, while the earlier solar abundance tables of 
\citet{gs98}, obtained using 1D solar atmospheric model, provide better agreement. Recent measurements of iron opacity in a 
condition similar to the solar interior shows that the iron opacity used in stellar models are significantly low 
\citep[see,][]{bail15}. This may partly resolve the issue of solar abundance problem, but not completely. It would be 
interesting to see if the asteroseismic data can tell us whether the problem is with opacities, or solar abundances, or with
both \citep[see, e.g.,][]{mazu10}. The position of the base of the convection zone is very sensitive to the opacity of the 
stellar material, which also depends on the abundance of the heavy elements, and can throw some light on the above issues.

\citet{lund16} have recently determined the oscillation frequencies of 66 main-sequence stars for which there are more than one 
year of {\it Kepler} data. This sample is known as {\it Kepler} seismic LEGACY sample. \citet{silv16} have used these set of 
frequencies to derive the properties of all stars in the sample. They found the masses, radii, and the ages with average 
uncertainties of about 4\%, 2\%, and 10\%, respectively. The long duration of the observations ensures sufficient precision to 
study the acoustic glitches in these stars. In this work, we use the above set of oscillation frequencies for all the 66 stars 
to estimate the acoustic depths of the base of the convection zone and helium ionization zone. We extend the work of 
\citet{silv16} by including the additional observables obtained from the glitch analysis to our stellar model fitting. 

The rest of the paper is organized as follows: Section~2 describes the spectroscopic and seismic data used in the study, 
the techniques to fit the glitch signature are described in Section~\ref{sec:fit}, 
Section~\ref{sec:model} presents the procedure to get the best-fit model, the results of the glitch analysis and stellar model 
fitting are discussed in Section~\ref{sec:results}, Section~\ref{sec:importance} demonstrates the importance of the glitch 
analysis in stellar model fitting, and finally we summarize the conclusions of this study in Section~\ref{sec:conc}.

\section{Spectroscopic and seismic data}
\begin{figure}
\plotone{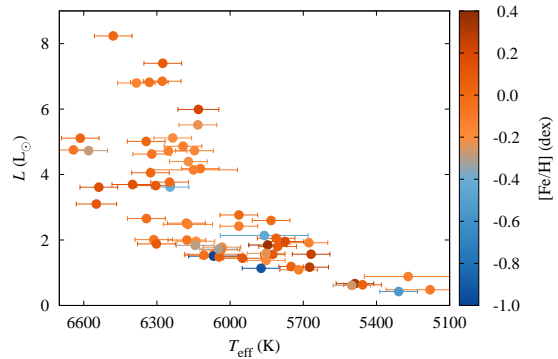}
\caption{Hertzsprung-Russell diagram showing the parameter space covered by the LEGACY sample. The luminosity of stars was 
obtained from the best-fit model. The colors represent the observed surface metallicity of the star.\label{fig1}}
\end{figure}

NASA's {\it Kepler} space mission has observed solar-like oscillations in over hundred Sun-like main-sequence stars. The 
LEGACY sample consists of 66 main-sequence stars observed in short cadence mode for at least 12 months (most of the stars have 
a time series of approximately 3 years), and spans a large range in metallicity. Figure~\ref{fig1} shows the locations of the 
stars in the Hertzsprung-Russell diagram. We used the spectroscopic and seismic data from \citet{lund16}, and refer the reader 
to that paper for the details on the target selection, compilation of the spectroscopic data, and the computation of the stellar 
oscillation frequencies \citep[see also,][]{silv16}. 

\section{Fitting techniques}
\label{sec:fit}
Acoustic glitches inside a star are regions where the sound speed or its derivatives show an abrupt variation on length scales 
shorter than the typical wavelengths of the acoustic modes. Such glitches introduce an oscillatory component, 
$\delta\nu_{\rm g}$, in the frequencies of stellar oscillations as a function of the radial order, $n$, of the form, 
$\delta\nu_{\rm g}(\nu) \propto \sin(4\pi\tau_{\rm g}\nu + \psi_{\rm g})$, where $\tau_{\rm g}$ is the acoustic depth of the 
glitch \citep[see,][]{goug88,voro88,goug90}, and $\psi_g$ is the phase of the
oscillatory signal.

The two main sources of acoustic glitches in a Sun-like main-sequence star are the base of the envelope convection zone (CZ) 
where the second derivative of the sound speed, $c$, is discontinuous, and the helium (He) ionization zone where the first 
adiabatic index, $\Gamma_1$, varies rapidly. Both of these glitches lie deep inside the star where the non-adiabatic 
effects are weak, and the glitch signatures are not significantly affected by the poorly understood near-surface layers. 
The boundary of the core convection zone does not contribute a significant 
oscillatory signal as this is aliased to a signal with a very small acoustic depth \citep{mazu01}, and that cannot be 
distinguished from the background smooth component of the frequency. It was customarily believed that the oscillatory signal 
from the helium ionization zone arises from the dip in $\Gamma_1$-profile in the second helium ionization zone. The fitted 
acoustic depth of the helium signal was, however, found to be significantly smaller than the depth of the He \textsc{ii} 
ionization zone, and \citet{houd07} attributed this difference to the neglect of the acoustic cut-off frequency in the phase 
function and the signal from the He \textsc{i} ionization zone. \citet{broo14} found that the fitted acoustic depth of the helium
signal in the models of the red-giants agrees with the acoustic depth of the peak in $\Gamma_1$-profile between the He \textsc{i}
and He \textsc{ii} ionization zones. \citet{verm14b} did a detailed study of glitch signals from various ionization zones of 
helium in main-sequence stellar models to find that the fitted acoustic depth always matches that of the peak in $\Gamma_1$ 
between the two ionization zones. Their attempt to fit the signatures from both ionization zones of helium resulted in a 
significantly better fit for the solar oscillation frequencies, but the fit was again to the usual peak between the two helium 
ionization zones, and a peak above the hydrogen ionization zone. More significantly, the inclusion of the second glitch did not 
affect the parameters for the helium glitch (the glitch between the two helium ionization zones). Hence in this work, we fit 
the glitch signature from the helium ionization zone only.

The amplitudes of the oscillatory signature from the acoustic glitches are approximately three or more orders of magnitude 
smaller than the background smooth component, which makes it hard to extract them from the stellar oscillation frequencies. There
are two popular approaches to extract the glitch signatures: the first attempts to fit the oscillation frequencies directly, 
while the second tries to fit the second differences of the oscillation frequencies. We use both fitting methods, Method A and 
Method B as described below, to derive the glitch parameters. The two independent methods can be used to assess the associated
systematic uncertainties in the estimated glitch properties.

\subsection{Fitting frequencies directly (Method A)}
\label{subsec:freq}
There are again two different approaches to extract the glitch signatures from the stellar oscillation frequencies: the first 
removes the smooth component from the frequencies as a function of radial order, $n$, and fits the residual 
\citep[see, e.g.,][]{mont94,mont98,mont00}, while the second approach fits the smooth component and the glitch signals 
simultaneously \citep[see, e.g.,][]{verm14a,verm14b}. We have used the second approach in this work as described below.

We model the smooth component of the oscillation frequency using a $l$-dependent fourth degree polynomial in the radial order, 
$n$, where $l$ is the harmonic degree. The functional forms of the glitch signatures are adapted from \citet{houd07}. We fit the 
oscillation frequency, $\nu_{n,l}$, to the function,
\begin{eqnarray}
f(n,l) = P_l(n) &+& \frac{a_c}{\nu^2} \sin(4\pi\tau_{\rm CZ}\nu + \psi_{\rm CZ})\nonumber\\
                &+& a_h \nu e^{-c_{2}\nu^2} \sin(4\pi\tau_{\rm He}\nu + \psi_{\rm He}),
\label{metha}
\end{eqnarray}
where $P_l(n) = \sum_{i = 0}^{4} a_i(l) n^i$ is the contribution of the smooth component with $a_i(l)$ being the coefficients of 
the polynomial. The second term is the oscillatory contribution coming from the base of the convection zone with $a_c$ related to
the amplitude, $\tau_{\rm CZ}$ being the acoustic depth of the base of the convection zone, and $\psi_{\rm CZ}$ being the phase 
of the signal. The third term is the oscillatory contribution coming from the helium ionization zone with $a_h$ related to 
the amplitude, $c_2$ related to the width of $\Gamma_1$-peak between the He \textsc{i} and He \textsc{ii} ionization zones, 
$\tau_{\rm He}$ being the acoustic depth of the $\Gamma_1$-peak, and $\psi_{\rm He}$ being the phase. This function contains a 
total of 22 free parameters when fitting $l$ = 0, 1, and 2 modes ($5 \times 3 = 15$ polynomial coefficients $a_i(l)$, $a_c$, 
$\tau_{\rm CZ}$, $\psi_{\rm CZ}$, $a_h$, $c_2$, $\tau_{\rm He}$, $\psi_{\rm He}$).

To determine the parameters of Eq.~(\ref{metha}), we perform a regularized least-squares fit by minimizing the function,
\begin{equation}
\chi_{\rm g}^2 = \sum_{n,l}\left[\frac{\nu_{n,l}-f(n,l)}{\sigma_{n,l}}\right]^2 + 
         \lambda^2 \sum_{n,l}\left[\frac{d^3P_{l}(n)}{dn^3}\right]^2,
\label{chi2g}
\end{equation}
where $\sigma_{n,l}$ is the quoted uncertainty on the observed $\nu_{n,l}$ and $\lambda$ is the regularization parameter. Note 
that we have used a third derivative regularization instead of the second derivative used in \citet{verm14a,verm14b}. The third 
derivative regularization marginally improves the stability of the fit. The regularization parameter is determined in the same 
way as in \citet{verm14a} for the solar oscillation frequencies, and the same value is used for stars in the {\it Kepler}  
LEGACY sample. The cost function $\chi_{\rm g}^2$ defined in Eq.~(\ref{chi2g}) is a nonlinear function of the fitting parameters, 
hence the fit may not converge to the global minimum, particularly if the initial guess is not close enough. We search for the 
global minimum in a subspace of the parameter space by repeating the fitting process for 100 sets of randomly chosen 
initial guesses. The fit with minimum value of the standard chi-square (first term in Eq.~\ref{chi2g}) among 100 trials is 
accepted as the best fit. We generate 1,000 
realizations of the observed oscillation frequencies assuming the uncertainties on them are uncorrelated 
and normally distributed. We fit all the realizations to get the distributions of the fitted parameters. The median of the 
distribution is accepted as the parameter value while the $16^{\rm th}$ and $84^{\rm th}$ percentiles of the distribution give 
the negative and positive errorbars.

The oscillatory signature from the base of the convection zone is typically weak with the amplitude of the order of the errors
on the oscillation frequencies. Consequently, the distribution of the fitted parameters may have multiple peaks. We use only 
those realizations for which the fitted acoustic depth falls in the dominant peak to calculate the median and error estimates. 
The parameters associated with the CZ 
signature may not be correct due to the problem of aliasing \citep{mazu01}, in which case the fitted acoustic depth is found to 
be the complement of $\tau_{\rm CZ}$, i.e., $\tau = T_0 - \tau_{\rm CZ}$ (where $T_0$ is the acoustic radius of the star). In 
spite of the problem of aliasing, the glitch analysis is useful as it can restrict $\tau_{\rm CZ}$ from its infinite possible 
values to two numbers, viz., $\tau_{\rm CZ}$ and $T_0 - \tau_{\rm CZ}$. In some cases, particularly for relatively massive 
stars with large errorbars on the oscillation frequencies, there may not be any well defined peak in the distribution of the 
fitted parameters for the convection zone signal, as the values may be spread over a wide interval.

\subsection{Fitting second differences (Method B)}
\label{subsec:diff}
This is a well known method for extracting the signatures of the acoustic glitches from the observed stellar oscillation
frequencies \citep[see, e.g.,][]{goug90,basu94,basu04}. The oscillation frequencies follow closely the asymptotic expression of 
\citet{tass80}, which predominantly depends linearly on the radial order for a given degree. Hence taking the second difference 
of the oscillation frequency with respect to $n$,
\begin{equation}
\delta^2\nu_{n,l} := \nu_{n-1,l} - 2 \nu_{n,l} + \nu_{n+1,l},
\end{equation}
reduces the background smooth component significantly. This, however, complicates slightly the fitting procedure because the 
differences have correlated errorbars, and the covariance matrix has to be used in the definition of the chi-square to be 
minimized.

We fit the second differences of the oscillation frequencies to the following function adapted from \citet{houd07},
\begin{eqnarray}
\delta^2\nu = a_0 + a_1 \nu &+& \frac{b_0}{\nu^2} \sin(4 \pi \nu \tau_{\rm CZ} + \phi_{\rm CZ})\nonumber\\
                            &+& c_0 \nu e^{-c_2 \nu^2} \sin(4 \pi \nu \tau_{\rm He} + \phi_{\rm He}),
\label{methb}
\end{eqnarray}
where the first two terms take care of the residual smooth component left after the second differences are calculated; the 
third term is the oscillatory contribution coming from the base of the convection zone with $b_0$ related to the amplitude, 
$\tau_{\rm CZ}$ being the acoustic depth of the base of the convection zone, and $\phi_{\rm CZ}$ being the phase; the fourth
term is the oscillatory contribution coming from the helium ionization zone with $c_0$ related to the amplitude, $c_2$ related 
to the width of $\Gamma_1$-peak between the He \textsc{i} and He \textsc{ii} ionization zones, $\tau_{\rm He}$ being the 
acoustic depth of the $\Gamma_1$-peak, and $\phi_{\rm He}$ being the phase. The above amplitudes of the oscillatory signatures 
in the second differences may be converted to the corresponding amplitudes in the oscillation frequencies by dividing them with 
$4\sin^2(2\pi\tau_{\rm g}\langle\Delta\nu\rangle) $\citep[see,][]{basu94}, where $\langle\Delta\nu\rangle$ is the average 
large frequency separation.

We fit the second differences of the oscillation frequencies to the function defined in Eq.~(\ref{methb}) to determine the 
parameters $a_0$, $a_1$, $b_0$, $\tau_{\rm CZ}$, $\phi_{\rm CZ}$, $c_0$, $c_2$, $\tau_{\rm He}$, and $\phi_{\rm He}$. We again 
search for the global minimum as in Method A with 100 trials on the initial guesses. The parameter values and the associated 
errorbars are computed using the distribution of the parameters obtained by fitting 10,000 realizations of the observed 
oscillation frequencies. The CZ signature has the same limitations as those discussed in the context of Method A.

\section{Best-fit models}
\label{sec:model}
We modeled each star in three different ways using various methods and evolutionary codes. The approaches are briefly described 
below. 

\subsection{MESA models}
We used the Modules for Experiments in Stellar Astrophysics code \citep[MESA;][]{paxt11,paxt13} for stellar modeling. This code 
can be used with various input physics and data tables. We used the OPAL equation of state \citep{roge02}, Opacity Project (OP) 
high-temperature opacities \citep{badn05,seat05} supplemented with low-temperature opacities from \citet{ferg05}. The metallicity
mixtures from \citet{gs98} was used. We used reaction rates from NACRE \citep{angu99} for all reactions except 
$^{14}{\rm N}(p,\gamma)^{15}{\rm O}$ and $^{12}{\rm C}(\alpha,\gamma)^{16}{\rm O}$, for which updated reaction rates from
\citet{imbr05} and \citet{kunz02} were used. Convection was modeled using the standard mixing-length theory \citep{cox68}. An
exponential overshoot \citep{herw00} was included for stars with masses greater than 1.10 M$_\odot$. The diffusion of helium and 
heavy elements was incorporated for stars of masses less than 1.35 M$_\odot$ using the prescription of \citet{thou94}. For higher
mass stars, the diffusion prescription clearly overestimates the settling of helium and heavy elements in the envelope, and hence
was not used. The adiabatic oscillation frequencies were calculated using the Adiabatic Pulsation code \citep[ADIPLS;][]{jcd08}.

We constructed models independently for each star in the LEGACY sample on a mesh of stellar parameters---the mass $M$, initial 
helium abundance $Y_i$, initial metallicity $[{\rm Fe}/{\rm H}]_i$, mixing-length $\alpha_{\rm MLT}$, and the overshoot parameter
$\alpha_{\rm OV}$. We generated 1,000 to 2,000 randomly distributed mesh points for an individual star in a reasonable subspace 
of the parameter space (we start with a chosen subspace with 1,000 mesh points, and extend it uniformly if the best-fit model falls near the edge). The models 
corresponding to every mesh point were evolved until the track enters in a box formed by the $4\sigma$ uncertainties in the 
observed effective temperature $T_{\rm eff}$, surface metallicity $[{\rm Fe}/{\rm H}]$, and average large frequency separation 
$\langle\Delta\nu\rangle$. We fitted the surface corrected model frequencies \citep{kjel08} to the observed ones to break the 
degeneracy inside the box, and accept the best-fit model as a representative model of the concerned star. In this manner, 
we get an ensemble of approximately 1,000 to 2,000 representative models (depending on the total number of mesh points) for each
star. Note that the number of models in the ensemble is not exactly same as the number of mesh points, because not all the tracks
enter the box. 

We took two different approaches to get the best-fit model from the above ensemble: the first approach used the conventional 
spectroscopic and seismic data but did not use the glitch information (termed as `{\it SeismicFit1}'), while the second used all 
the information including from the glitches (termed as `{\it GlitchFit}'). These two approaches are used only for the MESA 
models. In {\it SeismicFit1}, we defined a cost function, 
\begin{equation}
\chi_{\rm seismic}^2 = \sum_{p} \left(\frac{p_{\rm mod} - p_{\rm obs}}{\sigma_p}\right)^2,
\label{chi2m1}
\end{equation}
where $p$ represents 6 observable quantities; the $T_{\rm eff}$, $[{\rm Fe}/{\rm H}]$, large frequency separation averaged over 
the radial modes $\langle\Delta\nu\rangle_0$, average two-point frequency ratio $\langle r_{02} \rangle$, and the five-point 
ratios $r_{01}(n_0)$ and $r_{01}(n_0+3)$ ($n_0$ is suitably chosen radial order, and the choice of $n_0+3$ is made to avoid the 
correlation among the observables). We refer the reader to \citet{roxb03} for the definition of the ratios. The 
$\chi_{\rm seismic}^2$ was minimized over the model ensemble to get the best-fit model, and the uncertainties on the fitted 
parameters were estimated from the envelope of the $\chi_{\rm seismic}^2$ ($\Delta\chi_{\rm seismic}^2 = 1$). The detailed 
results for the LEGACY sample obtained using this approach were already presented in 
\citet[][see sections and results relevant to `V\&A']{silv16}. Here, we only compare some of those results with the results 
obtained when using the additional information from the glitch analysis (see the next paragraph).

In {\it GlitchFit}, we incorporated the information coming from the glitch analysis to our fitting pipeline. For this purpose, 
we fitted the signatures of the acoustic glitches in the frequencies of all the models in the ensemble using Method A to find 
the various parameters associated with the base of the convection zone and helium ionization zone, and defined a cost function, 
\begin{equation}
\chi_{\rm glitch}^2 = \chi_{\rm seismic}^2 + \sum_{q} \left(\frac{q_{\rm mod} - q_{\rm obs}}{\sigma_q}\right)^2,
\label{chi2m2}
\end{equation}
where $q$ represents 3 observable quantities; amplitude of helium signature averaged over the frequency range $A_{\rm He}$, width
of the $\Gamma_1$-peak $\Delta_{\rm He} = \sqrt{c_2/8\pi^2}$ \citep{houd07}, and the acoustic depth of the $\Gamma_1$-peak. Note 
that there is no ambiguity in comparing the acoustic depths of the helium ionization zone as obtained by fitting the observed and
model frequencies. A model representing the star must have similar helium glitch as the star, and should leave similar signature
on the oscillation frequencies, and hence the fitted $\tau_{\rm He}$ for both must be close. The differences arise when we try to
associate the fitted acoustic depth to a layer in the helium ionization zone \citep[see, e.g.,][]{broo14,verm14b}. We did not 
include the parameters associated with the CZ signature in the definition of $\chi_{\rm glitch}^2$ for the reasons that we 
discussed earlier in Section~\ref{subsec:freq}. We minimized the $\chi_{\rm glitch}^2$ over the ensemble to get the best-fit 
model, and the uncertainties on the fitted parameters were estimated in the same way as in {\it SeismicFit1}. The MESA model in 
the subsequent sections would always refer to the best-fit model obtained in this manner, unless stated otherwise.

The above methods find the best-fit model in two steps. In the first step, we fix the evolutionary stage for a set of initial
conditions ($M$, $Y_i$, $[{\rm Fe}/{\rm H}]_i$, $\alpha_{\rm MLT}$, $\alpha_{\rm OV}$) using oscillation frequencies, and filter
out reasonable models of the star. The oscillation frequencies monotonically decrease as a star evolves, and hence reasonably 
constrain the evolutionary stage for a given initial condition. The model frequencies have systematic uncertainties due to the
surface effect, hence we do not use them in the second step, instead use the quantities that are relatively insensitive to the 
surface effect, viz., the frequency ratios. In this manner, we use both the oscillation frequencies and their combinations to 
get the best-fit model. Since the methods preserve a set of reasonable models of the star, we may plot chi-square as a function 
of different stellar parameters, which gives additional useful information, e.g., the possibility of the secondary solutions.

\subsection{YREC models}
A second set of best-fit models were calculated using the Yale Rotating Stellar Evolution Code \citep[YREC;][]{dema08}. The 
models used OPAL high temperature opacities \citep{igle96} supplemented with low-temperature opacities of \citet{ferg05}. We used
the 2005 version of the OPAL equation of state \citep{roge02}. All nuclear reaction rates were from \citet{adel98} except for 
$^{14}$N($p$,$\gamma$)$^{15}O$ reaction, for which we used the updated rates of \citet{form04}. A large subset of the models 
included the diffusion of helium and other heavy elements with the diffusion coefficients from \citet{thou94}. The coefficients,
however, were changed with a multiplicative factor that depended on the mass of the models to inhibit the complete depletion of 
helium and heavy elements in the envelope convection zone. The coefficients were unchanged for the models with masses up to of 
1.25 M$_\odot$, while for higher masses the coefficients were multiplied by a factor, 
$f = \exp\left[-\frac{(M-1.25)^2}{2(0.085)^2}\right]$, where $M$ is the mass in solar unit.

The Yale Monte-Carlo Method \citep[YMCM;][]{silv15} was used to determine the best-fit model (`{\it SeismicFit2}').
This fitting method also does not use the glitch information and have been applied only to the YREC models. The reason for 
using two different names ({\it SeismicFit1} and {\it SeismicFit2}) is that the detailed optimization process and the observables 
used are different in the two cases. For each 
star, we start with using the average large frequency separation and frequency of maximum power along with the spectroscopic 
estimate of the effective temperature to get an estimate of the mass ($M$) and radius ($R$) of the star using the Yale Birmingham 
Grid-Based modeling pipeline \citep{basu10,gai11}. Since each of the observables has an associated error, we created several 
realizations of $M$, $R$, $T_{\rm eff}$, and $[{\rm Fe}/{\rm H}]$. For each realization ($M$, $R$, $T_{\rm eff}$, 
$[{\rm Fe}/{\rm H}]$), we used YREC in an iterative mode to obtain a model of the given mass $M$ and $[{\rm Fe}/{\rm H}]$ that 
had the required $R$ and $T_{\rm eff}$. This was done in two different ways: in the first approach, we kept $\alpha_{\rm MLT}$ 
fixed at different values and iterated over $Y_i$ to get the model; and in the second approach, we kept $Y_i$ fixed at different 
values and varied $\alpha_{\rm MLT}$ to get the required model.

\begin{figure*}
\epsscale{0.89}
\plotone{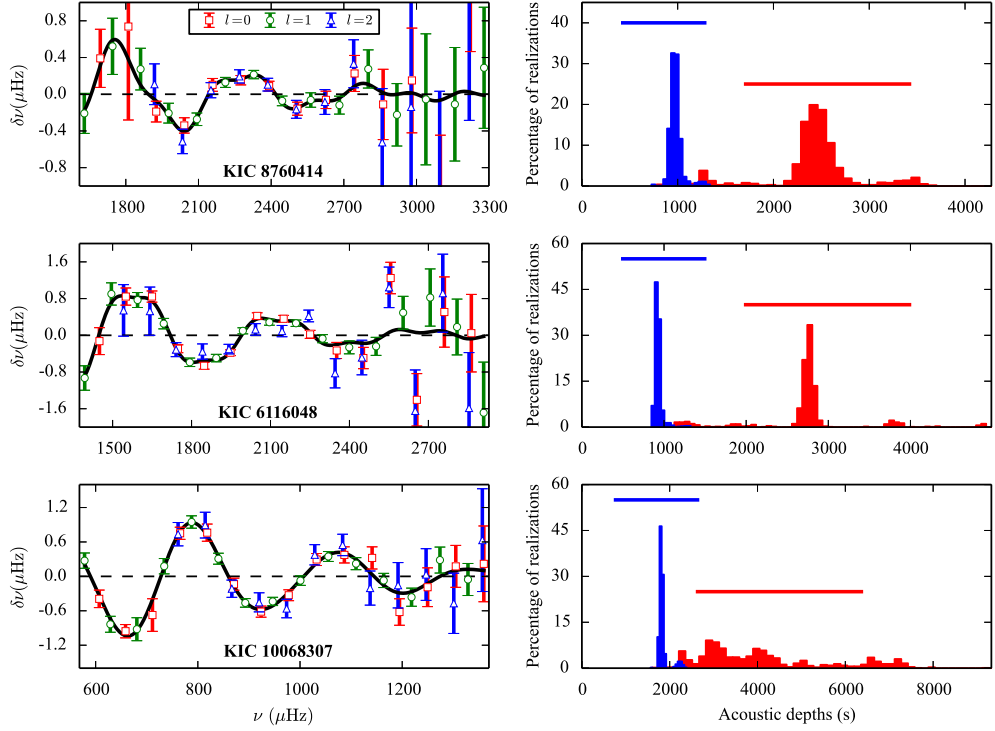}
\caption{Fits to the observed oscillation frequencies of three stars using Method A. The left panels show the oscillatory part 
of the frequency, $\delta\nu$, obtained by subtracting the smooth part, $P_l$, from them. The different points  
represent the observed frequencies, and the continuous line shows the best-fit to them. The right panels show the distribution 
of the acoustic depths of the CZ glitch (red histogram) and the He glitch (blue histogram). The horizontal bars show the ranges 
of initial guesses used for these two fitting parameters to search for the global minimum.\label{fig2}}
\end{figure*}

\begin{figure*}
\epsscale{0.89}
\plotone{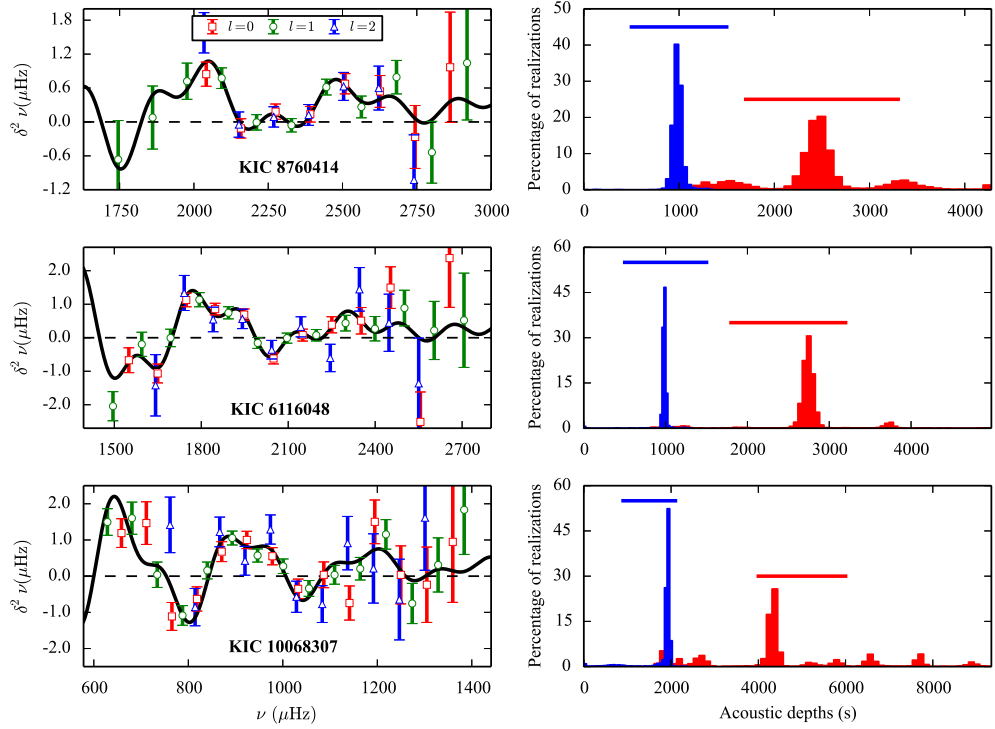}
\caption{Fits to the second differences of the observed oscillation frequencies using Method B. The different points in the left 
panels represent the observed second differences, and the continuous line shows the best-fit to them. The right panels show the 
distribution of the acoustic depths of the CZ glitch (red histogram) and the He glitch (blue histogram). The horizontal bars show
the ranges of initial guesses used for these two fitting parameters to search for the global minimum.\label{fig3}}
\end{figure*}

We computed the oscillation frequencies for all the models, and defined a cost function,
\begin{equation}
\chi^2_{\rm total} = \chi^2_{\nu} + \chi^2_{\rm ratios} + \chi^2_{T_{\rm eff}} + \chi^2_{\rm [Fe/H]},
\end{equation}
where $\chi^2_{\nu}$ was calculated using the surface corrected model frequencies \citep[two term formulation of][]{ball14}, 
while $\chi^2_{\rm ratios}$ was obtained using the uncorrected model frequencies (contains terms corresponding to both, $r_{02}$ 
and $r_{01}$). The first two terms on the right hand side are reduced chi-squares. Since the ratios are strongly correlated,
the full error covariance matrix was used to define $\chi^2_{\rm ratios}$. The best-fit model for a star was the one with the 
lowest value of $\chi^2_{\rm total}$.

\section{Results}
\label{sec:results}
We fitted the signatures of the acoustic glitches in the oscillation frequencies of all the 66 stars using both methods described
in Section~\ref{sec:fit}. The quality of fit primarily depends on the mass of star. Figures~\ref{fig2} and \ref{fig3} show 
respectively the fits obtained using Methods A and B for KIC 8760414, 6116048, and 10068307. These stars with masses close to 
0.80, 1.05, and 1.36 M$_\odot$, respectively, were selected to be representative of sub-solar, near-solar, and super-solar 
mass stars. Note the small amplitude of the helium signature in the fit for KIC 8760414. This is expected for the low-mass 
stars because the depression in their $\Gamma_1$-profile in the second helium ionization zone is shallow 
\citep[see, e.g.,][]{verm14b}, hence the amplitude of the peak between the He \textsc{i} and He \textsc{ii} ionization zones is 
small, consequently the amplitude of the helium signature is small. For low-mass stars, the small amplitude makes it difficult 
to fit the He signature unless sufficiently low radial order modes are observed. We found that the fit to the He signature was 
robust for all stars in the LEGACY sample, giving rise to a sharply peaked unimodal distribution of $\tau_{\rm He}$ 
(see, Figures~\ref{fig2} and \ref{fig3}). The fit to the CZ signature was also generally robust for stars of sub-solar and solar 
masses, with only a few problematic cases. However, fitting CZ signature was difficult for super-solar mass stars, particularly 
for stars of $M > 1.20$ M$_\odot$, which gave rise to multiple peaks in the distribution of $\tau_{\rm CZ}$. Such
stars are generally hot, and the envelope convection excites modes with shorter life-time, which leads to larger line-width 
of the modes and larger errorbar on the mode frequencies. For some problematic stars, most of the oscillation frequencies have 
errorbars that are larger than the average amplitude of the CZ signature.

We modeled each star in the LEGACY sample using the approaches described in Section~\ref{sec:model} to find the best-fit model 
and corresponding oscillation frequencies. Recall that the approach {\it GlitchFit} involves fitting the signatures of the 
acoustic glitches in the model frequencies using Method A. We used the same set of modes for the models as used for the 
observations. Table~\ref{tab1} lists the acoustic depths of the CZ and He glitch and average amplitude of the He glitch for 
both the observed frequencies and best-fit model frequencies from {\it GlitchFit}, as well as the mass, radius, and the age 
for all stars. 

\begin{table*}
\tiny
\tabletypesize{\tiny}
\begin{center}
\caption{Physical parameters for all stars in the {\it Kepler} seismic LEGACY sample.\label{tab1}}
\begin{tabular}{lrcccrrcccr}
\tableline
\tableline
& \multicolumn{6}{c}{Glitch parameters using Method A} & & \multicolumn{3}{c}{Stellar parameters using {\it GlitchFit}}\\
\cline{2-7} \cline{9-11}
KIC & $\tau_{\rm CZ,obs}$ & $\tau_{\rm CZ,mod}$ & $A_{\rm He,obs}$ & $A_{\rm He,mod}$ & $\tau_{\rm He,obs}$ & 
$\tau_{\rm He,mod}$ & & $M$ & $R$ & $t$\phantom{aaa} \\
& (s) & (s) & ($\mu$Hz) & ($\mu$Hz) & (s) & (s) & & (M$_\odot$) & (R$_\odot$) & (Gyr)\phantom{a}\\
\tableline
 1435467\tablenotemark{a} & $5307_{- 91}^{+ 80}$ & 2898 & $1.067_{-0.072}^{+0.076}$ & 0.957 & $1112_{- 35}^{+ 37}$ & 1056 & & $1.39\pm0.04$ & $1.706\pm0.020$ &  $2.7\pm0.3$\\ 
 2837475\tablenotemark{a} & $2422_{- 65}^{+ 65}$ & 1884 & $1.895_{-0.145}^{+0.144}$ & 1.609 & $ 905_{- 27}^{+ 26}$ &  843 & & $1.50\pm0.04$ & $1.659\pm0.020$ &  $1.7\pm0.2$\\
                  3427720 & $2068_{- 78}^{+ 99}$ & 2289 & $0.590_{-0.075}^{+0.079}$ & 0.633 & $ 817_{- 47}^{+ 56}$ &  668 & & $1.15\pm0.03$ & $1.130\pm0.015$ &  $2.4\pm0.2$\\
 3456181\tablenotemark{b} & $5259_{-504}^{+341}$ & 4373 & $1.088_{-0.122}^{+0.073}$ & 0.689 & $1689_{- 62}^{+ 56}$ & 1657 & & $1.50\pm0.04$ & $2.157\pm0.020$ &  $2.6\pm0.3$\\
                  3632418 & $4149_{-153}^{+188}$ & 4157 & $0.727_{-0.034}^{+0.030}$ & 0.655 & $1498_{- 35}^{+ 35}$ & 1420 & & $1.31\pm0.04$ & $1.867\pm0.020$ &  $2.8\pm0.3$\\
                  3656476 & $3710_{-126}^{+107}$ & 3575 & $0.536_{-0.055}^{+0.075}$ & 0.435 & $ 961_{- 32}^{+ 41}$ &  993 & & $1.09\pm0.03$ & $1.322\pm0.015$ &  $8.4\pm0.4$\\
                  3735871 & $1988_{- 75}^{+ 77}$ & 2323 & $0.598_{-0.071}^{+0.084}$ & 0.589 & $ 845_{- 70}^{+ 60}$ &  651 & & $1.20\pm0.04$ & $1.133\pm0.020$ &  $1.5\pm0.2$\\
                  4914923 & $3444_{-168}^{+258}$ & 3497 & $0.580_{-0.033}^{+0.032}$ & 0.553 & $1060_{- 32}^{+ 32}$ & 1025 & & $1.07\pm0.03$ & $1.357\pm0.015$ &  $6.8\pm0.3$\\
                  5184732 & $2810_{- 95}^{+ 70}$ & 3117 & $0.735_{-0.085}^{+0.086}$ & 0.663 & $ 849_{- 32}^{+ 38}$ &  826 & & $1.17\pm0.03$ & $1.329\pm0.015$ &  $4.0\pm0.4$\\
                  5773345 & $3352_{-143}^{+189}$ & 4099 & $0.798_{-0.065}^{+0.060}$ & 0.785 & $1505_{- 32}^{+ 38}$ & 1491 & & $1.55\pm0.04$ & $2.045\pm0.020$ &  $2.5\pm0.4$\\
                  5950854 & $3827_{-234}^{+191}$ & 3275 & $0.410_{-0.098}^{+0.170}$ & 0.236 & $1681_{-200}^{+164}$ & 1062 & & $1.03\pm0.02$ & $1.269\pm0.010$ & $10.3\pm0.4$\\
                  6106415 & $2707_{-111}^{+109}$ & 2685 & $0.542_{-0.033}^{+0.036}$ & 0.522 & $ 861_{- 30}^{+ 30}$ &  832 & & $1.05\pm0.04$ & $1.209\pm0.020$ &  $4.9\pm0.3$\\
                  6116048 & $2761_{- 60}^{+ 54}$ & 2790 & $0.467_{-0.027}^{+0.028}$ & 0.448 & $ 922_{- 23}^{+ 25}$ &  893 & & $1.05\pm0.02$ & $1.235\pm0.010$ &  $5.7\pm0.3$\\
                  6225718 & $2207_{-168}^{+328}$ & 2425 & $0.910_{-0.041}^{+0.042}$ & 0.919 & $ 773_{- 14}^{+ 15}$ &  736 & & $1.12\pm0.03$ & $1.220\pm0.015$ &  $2.5\pm0.2$\\
 6508366\tablenotemark{a} & $3325_{-161}^{+130}$ & 4688 & $0.936_{-0.077}^{+0.080}$ & 0.904 & $1527_{- 58}^{+ 61}$ & 1495 & & $1.37\pm0.04$ & $2.105\pm0.020$ &  $2.3\pm0.2$\\
                  6603624 & $3089_{- 68}^{+ 77}$ & 2976 & $0.370_{-0.023}^{+0.023}$ & 0.389 & $ 875_{- 28}^{+ 26}$ &  818 & & $1.05\pm0.02$ & $1.167\pm0.010$ &  $8.1\pm0.3$\\
                  6679371 & $3913_{-163}^{+182}$ & 2937 & $1.278_{-0.103}^{+0.094}$ & 1.200 & $1375_{- 54}^{+ 48}$ & 1352 & & $1.63\pm0.04$ & $2.248\pm0.020$ &  $2.0\pm0.3$\\
                  6933899 & $4395_{-206}^{+297}$ & 4179 & $0.466_{-0.024}^{+0.021}$ & 0.401 & $1499_{- 42}^{+ 41}$ & 1374 & & $1.12\pm0.03$ & $1.583\pm0.015$ &  $6.6\pm0.4$\\
 7103006\tablenotemark{a} & $2941_{-190}^{+268}$ & 3804 & $0.876_{-0.088}^{+0.100}$ & 0.644 & $1292_{- 77}^{+ 70}$ & 1407 & & $1.35\pm0.04$ & $1.903\pm0.020$ &  $2.4\pm0.3$\\
                  7106245 & $2916_{-234}^{+175}$ & 2236 & $0.542_{-0.120}^{+0.181}$ & 0.397 & $ 835_{-121}^{+115}$ &  793 & & $0.93\pm0.02$ & $1.100\pm0.010$ &  $7.1\pm0.3$\\
                  7206837 & $2665_{-107}^{+194}$ & 3097 & $1.113_{-0.115}^{+0.121}$ & 0.894 & $ 976_{- 40}^{+ 43}$ &  985 & & $1.36\pm0.04$ & $1.588\pm0.020$ &  $3.1\pm0.4$\\
                  7296438 & $3635_{-206}^{+210}$ & 3630 & $0.488_{-0.067}^{+0.068}$ & 0.463 & $1098_{- 81}^{+ 78}$ &  986 & & $1.15\pm0.03$ & $1.393\pm0.015$ &  $6.6\pm0.3$\\
 7510397\tablenotemark{a} & $4982_{-157}^{+155}$ & 3999 & $0.602_{-0.025}^{+0.028}$ & 0.535 & $1606_{- 39}^{+ 34}$ & 1469 & & $1.27\pm0.04$ & $1.821\pm0.020$ &  $3.3\pm0.3$\\
 7680114\tablenotemark{a} & $2697_{-257}^{+304}$ & 3739 & $0.435_{-0.041}^{+0.069}$ & 0.527 & $1205_{- 66}^{+ 61}$ & 1082 & & $1.13\pm0.03$ & $1.423\pm0.015$ &  $7.2\pm0.3$\\
                  7771282 & $3792_{-231}^{+229}$ & 3684 & $0.893_{-0.125}^{+0.137}$ & 0.776 & $1375_{-113}^{+104}$ & 1051 & & $1.30\pm0.03$ & $1.659\pm0.015$ &  $2.9\pm0.3$\\
                  7871531 & $2117_{- 59}^{+ 74}$ & 2091 & $0.237_{-0.049}^{+0.035}$ & 0.163 & $ 749_{- 84}^{+ 59}$ &  629 & & $0.80\pm0.02$ & $0.858\pm0.010$ &  $9.2\pm0.4$\\
                  7940546 & $3673_{- 93}^{+ 95}$ & 4368 & $0.754_{-0.041}^{+0.037}$ & 0.651 & $1551_{- 38}^{+ 37}$ & 1463 & & $1.33\pm0.03$ & $1.915\pm0.015$ &  $2.7\pm0.3$\\
                  7970740 & $1955_{- 79}^{+ 65}$ & 1824 & $0.215_{-0.041}^{+0.045}$ & 0.198 & $ 551_{- 48}^{+ 38}$ &  580 & & $0.73\pm0.03$ & $0.761\pm0.015$ & $10.1\pm0.4$\\
                  8006161 & $2264_{- 83}^{+ 92}$ & 2163 & $0.424_{-0.045}^{+0.053}$ & 0.457 & $ 563_{- 20}^{+ 20}$ &  552 & & $1.00\pm0.03$ & $0.933\pm0.015$ &  $4.9\pm0.2$\\
 8150065\tablenotemark{a} & $3749_{-157}^{+154}$ & 2779 & $0.707_{-0.707}^{+0.506}$ & 0.629 & $1012_{-358}^{+371}$ &  854 & & $1.12\pm0.03$ & $1.367\pm0.015$ &  $4.0\pm0.3$\\
                  8179536 & $2363_{- 55}^{+ 61}$ & 2308 & $1.076_{-0.110}^{+0.115}$ & 1.000 & $ 803_{- 55}^{+ 55}$ &  757 & & $1.20\pm0.03$ & $1.331\pm0.015$ &  $1.6\pm0.2$\\
                  8228742 & $4521_{- 94}^{+ 93}$ & 4736 & $0.545_{-0.028}^{+0.031}$ & 0.509 & $1578_{- 38}^{+ 38}$ & 1491 & & $1.28\pm0.03$ & $1.825\pm0.015$ &  $4.6\pm0.4$\\
                  8379927 & $2219_{-209}^{+222}$ & 2160 & $0.704_{-0.033}^{+0.029}$ & 0.654 & $ 703_{- 22}^{+ 21}$ &  659 & & $1.11\pm0.03$ & $1.114\pm0.015$ &  $1.7\pm0.2$\\
 8394589\tablenotemark{b} & $1952_{-148}^{+108}$ & 2373 & $0.793_{-0.058}^{+0.065}$ & 0.683 & $ 750_{- 44}^{+ 50}$ &  769 & & $1.08\pm0.04$ & $1.178\pm0.020$ &  $3.9\pm0.3$\\
                  8424992 & $2662_{-256}^{+312}$ & 2636 & $0.412_{-0.116}^{+0.178}$ & 0.179 & $1045_{-210}^{+155}$ &  766 & & $0.95\pm0.03$ & $1.060\pm0.015$ &  $9.1\pm0.3$\\
                  8694723 & $3281_{- 60}^{+ 69}$ & 3793 & $0.693_{-0.038}^{+0.035}$ & 0.678 & $1245_{- 30}^{+ 31}$ & 1220 & & $1.09\pm0.03$ & $1.522\pm0.015$ &  $4.5\pm0.3$\\
                  8760414 & $2455_{-127}^{+131}$ & 2528 & $0.222_{-0.026}^{+0.029}$ & 0.181 & $ 968_{- 37}^{+ 36}$ &  929 & & $0.80\pm0.02$ & $1.018\pm0.010$ & $12.4\pm0.4$\\
                  8938364 & $4044_{- 69}^{+ 66}$ & 3891 & $0.390_{-0.023}^{+0.024}$ & 0.343 & $1243_{- 39}^{+ 45}$ & 1170 & & $1.06\pm0.03$ & $1.386\pm0.015$ &  $9.8\pm0.4$\\
 9025370\tablenotemark{b} & $1722_{-105}^{+108}$ & 2321 & $0.322_{-0.055}^{+0.103}$ & 0.244 & $ 810_{-144}^{+137}$ &  643 & & $1.02\pm0.03$ & $1.021\pm0.015$ &  $5.0\pm0.3$\\
                  9098294 & $3128_{-213}^{+133}$ & 2833 & $0.449_{-0.046}^{+0.041}$ & 0.343 & $ 837_{- 26}^{+ 30}$ &  856 & & $0.97\pm0.02$ & $1.145\pm0.010$ &  $8.0\pm0.3$\\
 9139151\tablenotemark{a} & $3239_{- 79}^{+ 88}$ & 2277 & $0.683_{-0.062}^{+0.063}$ & 0.698 & $ 788_{- 33}^{+ 30}$ &  685 & & $1.20\pm0.03$ & $1.168\pm0.015$ &  $1.9\pm0.2$\\
 9139163\tablenotemark{a} & $2182_{- 51}^{+ 44}$ & 2318 & $1.335_{-0.067}^{+0.073}$ & 1.138 & $ 865_{- 22}^{+ 22}$ &  889 & & $1.34\pm0.04$ & $1.542\pm0.020$ &  $2.2\pm0.3$\\
                  9206432 & $2839_{- 62}^{+ 64}$ & 2351 & $1.489_{-0.100}^{+0.117}$ & 1.005 & $ 892_{- 34}^{+ 39}$ &  845 & & $1.38\pm0.04$ & $1.513\pm0.020$ &  $2.0\pm0.3$\\
 9353712\tablenotemark{b} & $6305_{-165}^{+154}$ & 4028 & $0.775_{-0.097}^{+0.103}$ & 0.766 & $1612_{- 69}^{+ 71}$ & 1599 & & $1.53\pm0.04$ & $2.178\pm0.020$ &  $2.7\pm0.3$\\
                  9410862 & $2760_{-106}^{+125}$ & 2692 & $0.430_{-0.077}^{+0.055}$ & 0.374 & $1065_{- 53}^{+ 58}$ &  838 & & $1.03\pm0.03$ & $1.178\pm0.015$ &  $6.4\pm0.3$\\
                  9414417 & $4289_{-199}^{+157}$ & 4665 & $0.768_{-0.055}^{+0.058}$ & 0.694 & $1468_{- 38}^{+ 41}$ & 1396 & & $1.29\pm0.03$ & $1.865\pm0.015$ &  $3.5\pm0.3$\\
 9812850\tablenotemark{b} & $4671_{-196}^{+363}$ & 4368 & $1.154_{-0.103}^{+0.097}$ & 0.782 & $1291_{- 41}^{+ 44}$ & 1297 & & $1.26\pm0.04$ & $1.760\pm0.020$ &  $3.3\pm0.3$\\
                  9955598 & $1940_{-417}^{+115}$ & 2089 & $0.250_{-0.115}^{+0.627}$ & 0.195 & $ 980_{-572}^{+360}$ &  548 & & $0.87\pm0.02$ & $0.877\pm0.010$ &  $6.5\pm0.3$\\
 9965715\tablenotemark{b} & $2800_{-150}^{+270}$ & 2318 & $0.907_{-0.091}^{+0.111}$ & 0.843 & $ 825_{- 53}^{+ 45}$ &  738 & & $1.07\pm0.04$ & $1.269\pm0.020$ &  $3.2\pm0.3$\\
10068307\tablenotemark{b} & $4000_{-318}^{+298}$ & 5349 & $0.601_{-0.036}^{+0.031}$ & 0.508 & $1813_{- 31}^{+ 27}$ & 1719 & & $1.36\pm0.03$ & $2.053\pm0.015$ &  $3.2\pm0.4$\\
                 10079226 & $2960_{-372}^{+101}$ & 2460 & $0.518_{-0.088}^{+0.109}$ & 0.494 & $ 870_{-129}^{+105}$ &  678 & & $1.11\pm0.03$ & $1.145\pm0.015$ &  $2.8\pm0.2$\\
                 10162436 & $4316_{-253}^{+211}$ & 4024 & $0.775_{-0.040}^{+0.037}$ & 0.767 & $1579_{- 37}^{+ 39}$ & 1499 & & $1.18\pm0.04$ & $1.898\pm0.020$ &  $3.7\pm0.3$\\
                 10454113 & $2189_{- 94}^{+ 74}$ & 2362 & $1.079_{-0.075}^{+0.071}$ & 0.942 & $ 734_{- 22}^{+ 19}$ &  722 & & $1.16\pm0.03$ & $1.237\pm0.015$ &  $1.9\pm0.2$\\
                 10516096 & $3646_{-102}^{+120}$ & 3602 & $0.484_{-0.036}^{+0.034}$ & 0.432 & $1167_{- 42}^{+ 41}$ & 1112 & & $1.17\pm0.03$ & $1.447\pm0.015$ &  $6.2\pm0.3$\\
10644253\tablenotemark{b} & $1804_{- 82}^{+ 65}$ & 2229 & $0.866_{-0.091}^{+0.103}$ & 0.907 & $ 675_{- 42}^{+ 36}$ &  616 & & $1.12\pm0.03$ & $1.105\pm0.015$ &  $1.1\pm0.1$\\
                 10730618 & $3691_{- 89}^{+100}$ & 3947 & $1.207_{-0.138}^{+0.132}$ & 0.701 & $1260_{- 47}^{+ 54}$ & 1235 & & $1.23\pm0.04$ & $1.729\pm0.020$ &  $3.1\pm0.3$\\
                 10963065 & $2900_{-190}^{+180}$ & 2716 & $0.659_{-0.055}^{+0.065}$ & 0.646 & $ 793_{- 40}^{+ 47}$ &  826 & & $1.08\pm0.03$ & $1.228\pm0.015$ &  $4.1\pm0.3$\\
                 11081729 & $2339_{-325}^{+206}$ & 2802 & $1.022_{-0.139}^{+0.155}$ & 0.994 & $ 923_{-121}^{+ 99}$ &  740 & & $1.25\pm0.04$ & $1.416\pm0.020$ &  $2.2\pm0.3$\\
                 11253226 & $1954_{- 73}^{+ 95}$ & 2018 & $2.016_{-0.119}^{+0.105}$ & 1.965 & $ 843_{- 22}^{+ 23}$ &  774 & & $1.43\pm0.04$ & $1.610\pm0.020$ &  $1.9\pm0.3$\\
                 11772920 & $1847_{-170}^{+368}$ & 2042 & $0.337_{-0.091}^{+0.131}$ & 0.152 & $ 524_{- 92}^{+ 83}$ &  573 & & $0.78\pm0.03$ & $0.829\pm0.015$ &  $9.5\pm0.5$\\
                 12009504 & $2996_{-108}^{+ 94}$ & 2992 & $0.722_{-0.058}^{+0.058}$ & 0.687 & $ 938_{- 36}^{+ 50}$ &  885 & & $1.14\pm0.03$ & $1.391\pm0.015$ &  $3.6\pm0.3$\\
12069127\tablenotemark{a} & $3902_{-113}^{+108}$ & 4661 & $0.863_{-0.128}^{+0.094}$ & 0.733 & $1745_{-107}^{+113}$ & 1697 & & $1.61\pm0.06$ & $2.314\pm0.030$ &  $2.5\pm0.3$\\
                 12069424 & $2839_{- 73}^{+ 73}$ & 3063 & $0.451_{-0.037}^{+0.044}$ & 0.418 & $ 910_{- 32}^{+ 31}$ &  886 & & $1.10\pm0.02$ & $1.237\pm0.010$ &  $6.7\pm0.3$\\
12069449\tablenotemark{b} & $1790_{- 60}^{+ 55}$ & 2724 & $0.480_{-0.023}^{+0.023}$ & 0.457 & $ 767_{- 17}^{+ 20}$ &  768 & & $1.00\pm0.03$ & $1.102\pm0.015$ &  $6.9\pm0.3$\\
                 12258514 & $3906_{-163}^{+218}$ & 3864 & $0.548_{-0.021}^{+0.018}$ & 0.526 & $1274_{- 20}^{+ 23}$ & 1195 & & $1.22\pm0.03$ & $1.586\pm0.015$ &  $4.3\pm0.3$\\
12317678\tablenotemark{a} & $2545_{- 72}^{+ 68}$ & 3991 & $1.498_{-0.083}^{+0.089}$ & 1.035 & $1115_{- 30}^{+ 32}$ & 1260 & & $1.19\pm0.04$ & $1.758\pm0.020$ &  $3.4\pm0.3$\\
\tableline
\end{tabular}
\tablecomments{The symbols in the header have usual meaning. The quantities with subscript `obs' are obtained by fitting the 
observed frequencies, while with `mod' are found by fitting the best-fit model frequencies from {\it GlitchFit}.}
\tablenotetext{1}{The distribution of $\tau_{\rm CZ}$ have multiple peaks. There is a small peak close to the $\tau_{\rm CZ}$
obtained using sound-speed profile of the best-fit model, but the acoustic depth corresponding to the dominant peak and its 
complement are far from the $\tau_{\rm CZ}$. This may happen if the signature is not significant, i.e., the oscillation 
frequencies have large errorbars.}
\tablenotetext{2}{The fitted acoustic depth of the base of the convection zone is close to the complement of the $\tau_{\rm CZ}$
obtained using sound-speed profile of the best-fit model.}
\end{center}
\end{table*}

\begin{figure*}
\plotone{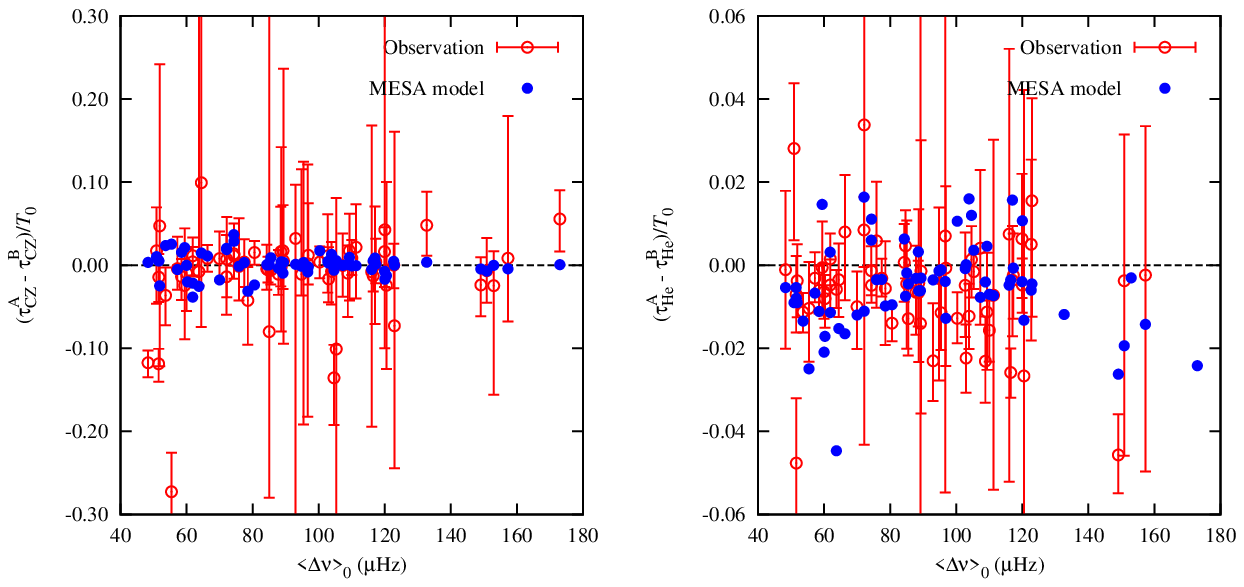}
\caption{Comparison of the acoustic depths (base of the convection zone in the left panel and helium ionization zone in the right
panel) obtained using Methods A and B for all stars in the LEGACY sample. The open circles with errorbar correspond to the fit to
the observed frequencies, while filled circles to the fit to the best-fit model frequencies.\label{fig4}}
\end{figure*}

For the sake of a clear presentation, we show here that the results obtained using Methods A and B agree very well, and then 
present the results obtained using only Method A in most of the subsequent sections. Figure~\ref{fig4} shows the differences 
between the acoustic depths obtained using Methods A and B. We can see an excellent level of agreement between the results of 
the two methods. This, however, does not guarantee the accuracy of the results. In fact, the acoustic depth of the base of the 
convection zone is incorrect for some stars in the sample. In such cases, both methods give systematically incorrect 
$\tau_{\rm CZ}$, as also noted by \citet{rees16}.

\subsection{Acoustic depths of the CZ and He glitches}
The glitches observed in Sun-like main-sequence stars are regions of only sharp variation in sound speed (not discontinuities
in sound speed), and are extended in depth, particularly the glitch arising from the helium ionization zone. To compare the 
acoustic depths obtained using glitch analysis with the acoustic depth of the layer which causes the signature, we computed 
the acoustic depths of layers using sound-speed profile of the best-fit model,
\begin{equation}
\tau_r = \int_{r}^{R_*} \frac{dr}{c}, 
\end{equation}
where $r$ is the radial distance of the layer, $R_*$ the radius of the star (to the acoustic surface and not to the photosphere), 
and $c$ is the sound speed.

\begin{figure*}
\plotone{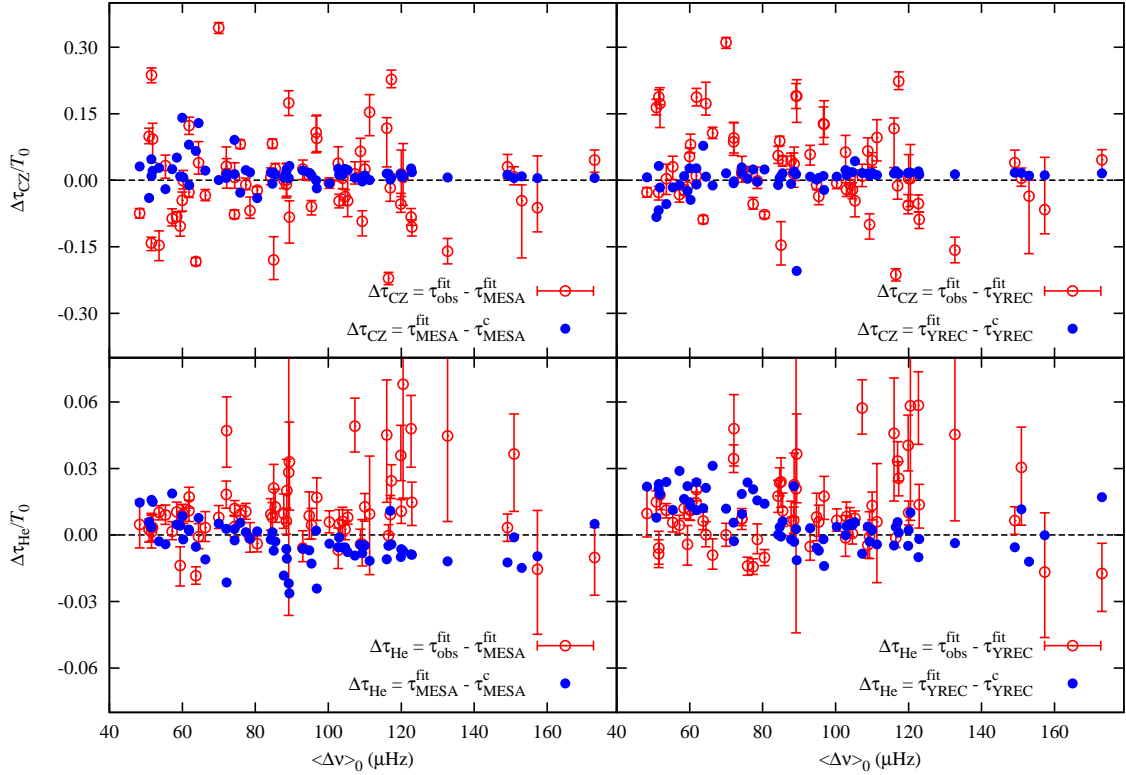}
\caption{Comparison of the acoustic depths obtained using glitch analysis (Method A) and the sound-speed profiles (MESA/YREC). 
The open circles with errorbar show the scaled difference between the acoustic depths obtained by fitting the observed 
($\tau_{\rm obs}^{\rm fit}$) and best-fit model ($\tau_{\rm MESA/YREC}^{\rm fit}$) frequencies, while the blue circles show the 
scaled difference between the acoustic depths obtained by fitting the best-fit model frequencies and using the sound-speed 
profile ($\tau_{\rm MESA/YREC}^{\rm c}$). The $\tau_{\rm MESA}^{\rm c}$ and $\tau_{\rm YREC}^{\rm c}$ in the bottom panels 
represent the acoustic depth of the peak in first adiabatic index between the He \textsc{i} and He \textsc{ii} ionization zones.
\label{fig5}}
\end{figure*}

The top panel of Figure~\ref{fig5} compares the different estimates of the acoustic depth of the base of the convection zone. 
The acoustic depth obtained by fitting the best-fit model frequencies agrees quite well with the acoustic depth calculated using 
the corresponding sound-speed profile. The calculation of the acoustic depth using sound-speed profile requires the definition 
of the acoustic surface, which is uncertain. \citet{balm90} have argued that the acoustic surface of a star should be defined at 
a radial distance in the atmosphere where the extrapolated $c^2$ from the outer convection zone vanishes 
\citep[see also,][]{lope01}. An uncertainty in the definition of the acoustic surface introduces a fixed shift in the acoustic 
depths calculated using sound-speed profile. Here, we assumed the acoustic surface to be the top most layer of the Eddington 
atmosphere ($\tau = 10^{-5}$). The scaled differences of less than 0.03 between the fitted and calculated acoustic depths 
suggest that the true acoustic surface is not very far from the assumed layer. The points corresponding to the difference 
between the acoustic depths obtained by fitting the observed and best-fit model frequencies are more scattered. This is 
primarily due to the fact that the observed frequencies have associated observational uncertainties, and the fit to the weak 
CZ signature in the observed frequencies is more prone to aliasing than the fit to the model frequencies. 

The helium ionization zones are extended in depth. Traditionally, it has been assumed while modeling the form of the He glitch 
signature that it arises from the He \textsc{ii} ionization zone \citep[see, e.g.,][]{mont98,houd07}, which implies that 
the fitted acoustic depth should represent a layer in the He \textsc{ii} ionization zone where $\Gamma_1$ is minimum (`dip'). 
Recently, \citet{broo14} and \citet{verm14b} found respectively in the red-giant and main-sequence stellar models that the fitted 
acoustic depth corresponds more closely to a layer between the He \textsc{i} and He \textsc{ii} ionization zones where $\Gamma_1$
is maximum (`peak'). The bottom panel of Figure~\ref{fig5} compares the different estimates of the acoustic depth of the helium 
ionization zone. The acoustic depths obtained by fitting the observed and best-fit model frequencies are in good agreement, as 
expected. We confirm that the fitted acoustic depth matches with the acoustic depth of the peak in $\Gamma_1$-profile. 

\begin{figure*}
\plotone{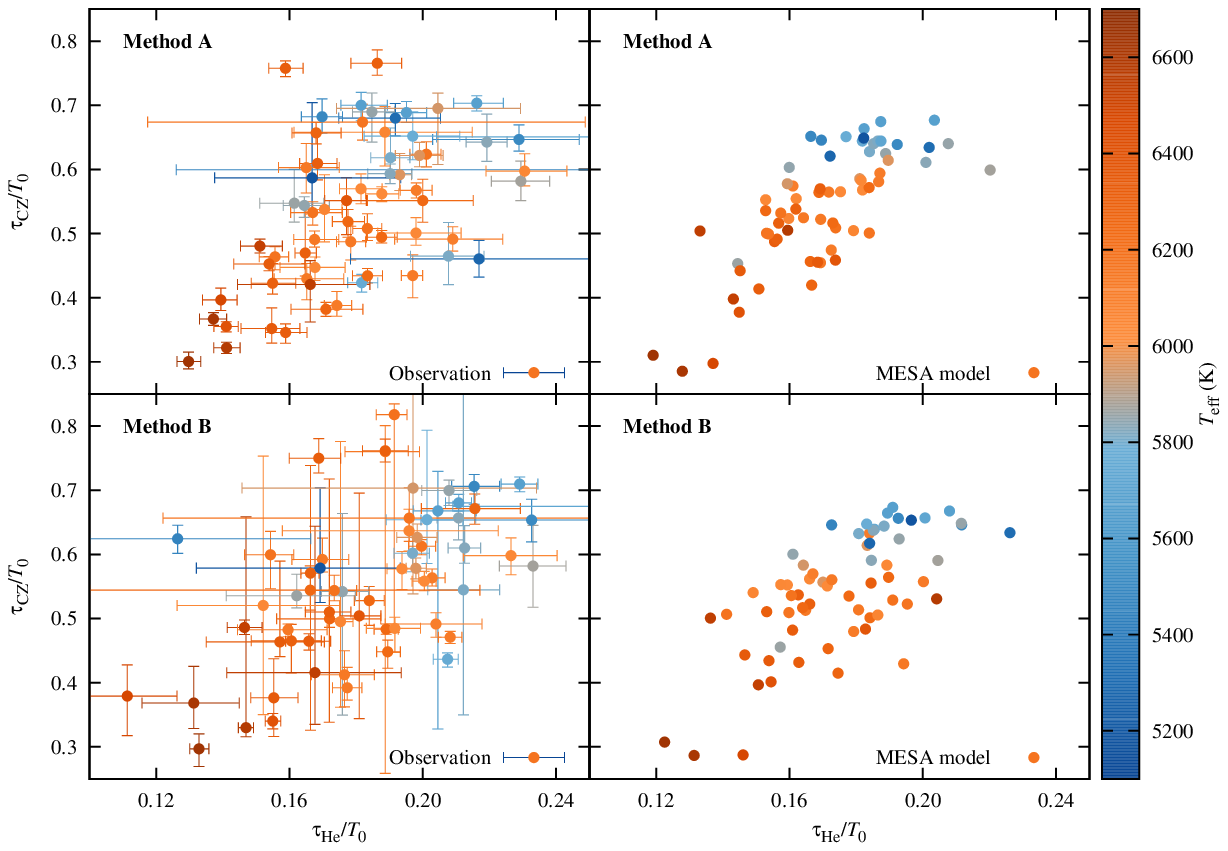}
\caption{Scaled acoustic depths of the base of the convection zone and helium ionization zone as obtained using glitch analysis. 
A point in a panel represents a star. The points with errorbar in the left panels were obtained by fitting the observed 
frequencies, while points in the right panels were found by fitting the best-fit model frequencies. The colors represent the 
effective temperature of the star.\label{fig6}}
\end{figure*}

\subsection{Ensemble study}
Figure~\ref{fig6} shows the acoustic depths of the CZ and He glitch for all the 66 stars in the sample. The results obtained
using the two methods look very similar, except that the errorbars obtained using Method B is on average larger than Method A, 
particularly the errorbar on the acoustic depth of the helium ionization zone. This may be expected because the errorbars on the 
second differences increase by about a factor 2.5 in comparison to the errorbars on the oscillation frequencies, while the 
amplitude of the signature increases approximately by a factor $4\sin^2(2\pi\tau_{\rm g}\langle\Delta\nu\rangle)$ \citep{basu94}. The 
factor $4\sin^2(2\pi\tau_{\rm g}\langle\Delta\nu\rangle)$ is generally smaller than 2.5 for the He signature ($\approx$ 1.8, 1.2, 
and 1.3 for KIC 8760414, 6116048, and 10068307, respectively), effectively reducing its significance in the second 
differences in comparison to the frequencies. There is a clear correlation seen in Figure~\ref{fig6} between the acoustic depths 
of the base of the convection zone and helium ionization zone. The larger scatter seen in the left panels is mostly due to 
aliasing of the CZ signature. The helium signature is typically strong, and the determination of $\tau_{\rm He}$ is reliable. 
The fitted $\tau_{\rm He}$ together with the above correlation may be used to select the correct solution in the cases where the 
distribution of $\tau_{\rm CZ}$ have multiple peaks. As one may expect, the figure also shows that the cooler stars have deeper 
convection zones as well as deeper helium ionization layers.

\begin{figure*}
\epsscale{0.99}
\plotone{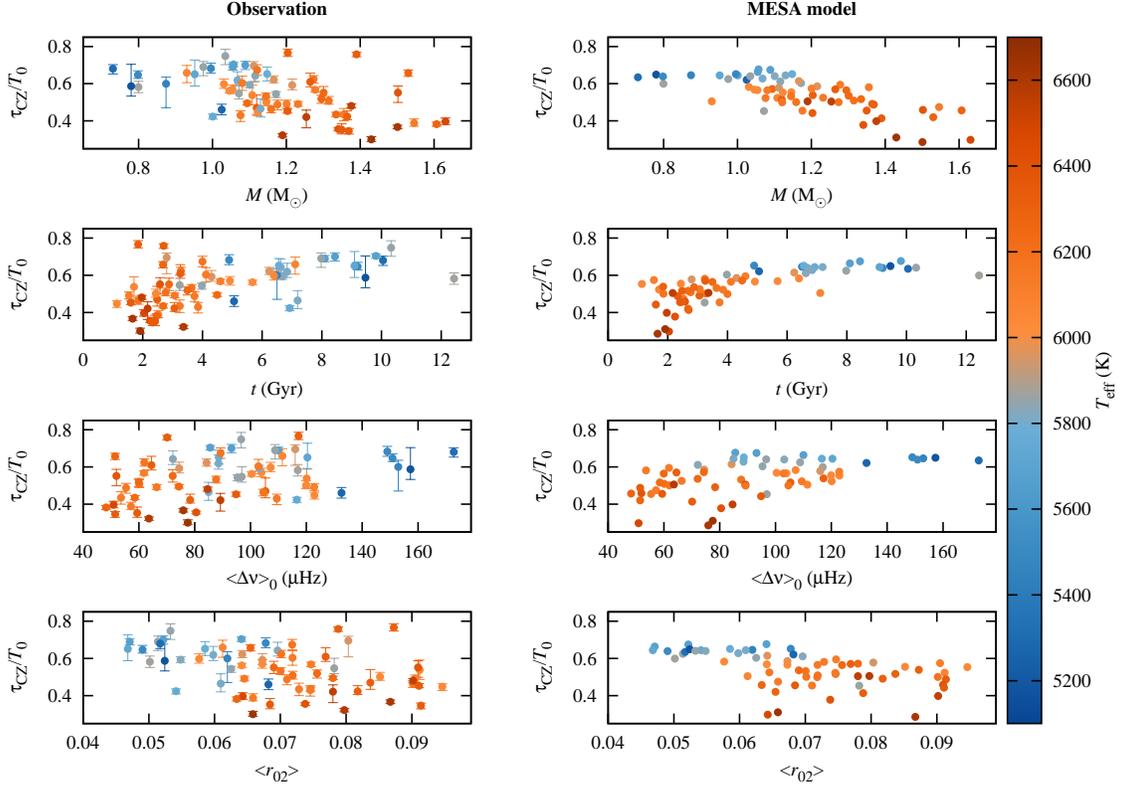}
\caption{Scaled acoustic depth of the base of the convection zone obtained using Method A as a function of different stellar
parameters. The points with errorbar in the left panels were obtained by fitting the observed frequencies, while points in the 
right panels were found by fitting the best-fit model frequencies. The colors represent the effective temperature of the star.
\label{fig7}}
\end{figure*}

Figure~\ref{fig7} shows the scaled acoustic depth of the base of the convection zone obtained using Method A as a function 
of mass, age, large frequency separation averaged over radial modes, and average two-point ratio. The fit to the model 
frequencies is not as much affected by the problem of aliasing as the observed frequencies, unveiling the relationships between 
the acoustic depth of the base of the convection zone and various stellar properties better. The results obtained using 
Method B look very similar. As can be seen from the topmost panels, the acoustic depth of the base of the convection zone 
decreases as the mass (hence the effective temperature) increases. This is expected as the hotter stars have smaller 
opacity, and the radiation can transport the energy in the larger part of the envelope. The acoustic depth of the base of the 
convection zone increases as a function of the age and average large frequency separation, while it decreases with the two-point 
ratio. The age and large frequency separation depend on the mass of the star, and the global trend seen in the corresponding 
panels are result of that dependence. The two-point ratio is an indicator of the evolutionary stage of the star---it decreases
as star evolves on the main-sequence---and the dependence seen in the bottom panel can be largely understood in terms of its
dependence on the age. The dependence of the $\tau_{\rm CZ}$ on composition is significantly weaker than on the mass, and can 
be seen only if the mass is constrained to a narrow range. 

\begin{figure*}
\epsscale{0.99}
\plotone{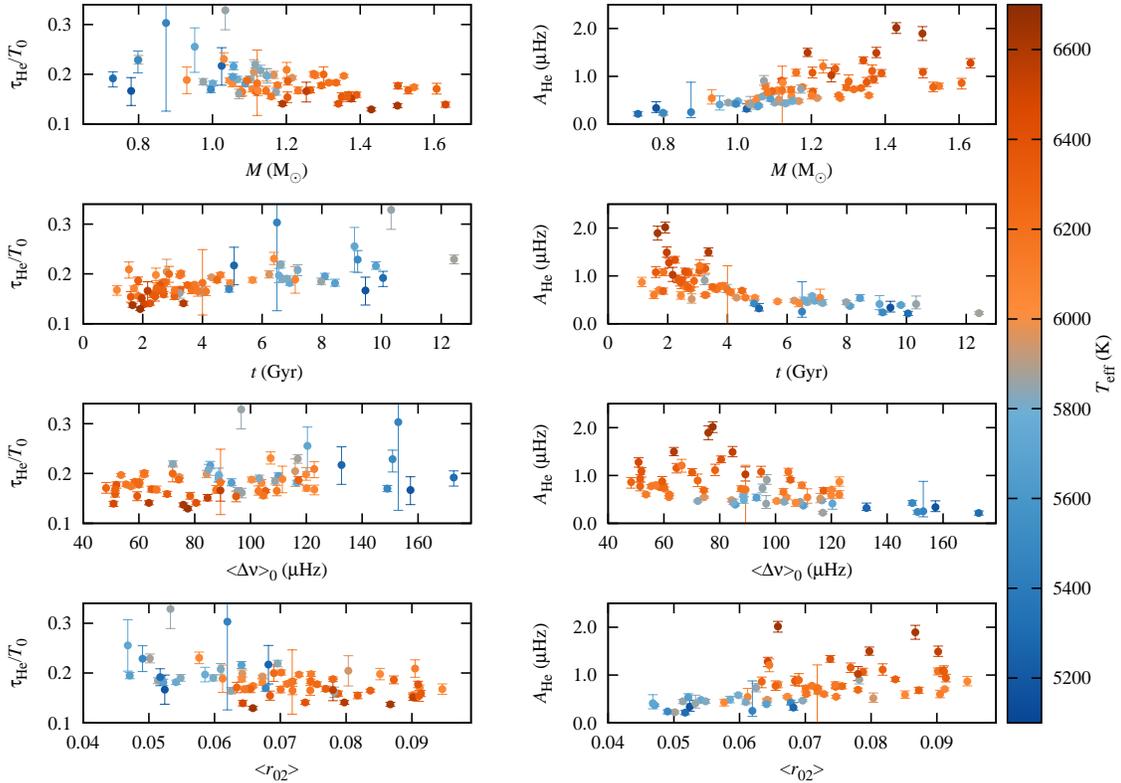}
\caption{Scaled acoustic depth of the helium ionization zone (left panels) and average amplitude of helium signature (right 
panels) as a function of different stellar parameters. The acoustic depth and the average amplitude were obtained by fitting the
observed oscillation frequencies using Method A. The colors represent the effective temperature of the star.\label{fig8}}
\end{figure*}

The helium signature in the oscillation frequencies of the Sun-like stars cannot only be used to derive the location of the 
helium ionization zone but also can be used to estimate the envelope helium abundance. The average amplitude of the helium 
signature depends on the amount of helium present in its ionization zone \citep[see, e.g.,][]{basu04,mont05,houd07}, which may 
be calibrated against the corresponding amplitudes of the helium signatures in the model frequencies of different envelope 
helium abundance to estimate its abundance \citep{verm14a}. Figure~\ref{fig8} shows the acoustic depth of the helium ionization 
zone as well as the average amplitude of the helium signature as a function of $M$, $t$, $\langle\Delta\nu\rangle_0$, and 
$r_{02}$. The acoustic depth of the helium ionization zone decreases as the mass increases. 
This is again expected as the helium gets ionized closer to the surface for hotter stars. The acoustic depth of the helium 
ionization zone increases as a function of the age and large frequency separation, while it decreases with the two-point ratio. 
The average amplitude of the helium signature increases with the mass, as was noted by \citet{verm14b}. This complicates the 
calibration involved in the helium abundance determination. The amplitude decreases as a function of the age and large frequency 
separation, while it increases with the two-point ratio. The variation of the amplitude and the acoustic depth with $t$, 
$\langle\Delta\nu\rangle_0$, and $r_{02}$ can again be understood mostly in terms of the variation of these parameters with 
the mass and effective temperature.

\section{The importance of analyzing acoustic glitches}
\label{sec:importance}
The stellar model fitting is a high-dimensional non-linear optimization problem, and the solution may not always converge to 
the global minimum. There are several fitting methods in use, e.g., parallel genetic algorithm \citep{metc09}, Bayesian 
approach \citep{grub12,silv15}, machine learning method \citep{verm16,bell16}, etc. The performance of different fitting methods 
have been compared in the past \citep[see, e.g.,][]{rees16,silv16}. The trouble common to all the methods is that the stellar 
parameters have intrinsic correlations, e.g., the well known anti-correlation between the mass and initial helium abundance 
\citep[see, e.g.,][]{metc09,lebr14,verm16}, and they are not well constrained by the conventional spectroscopic and seismic data,
particularly the initial helium abundance. 

\begin{figure*}
\plotone{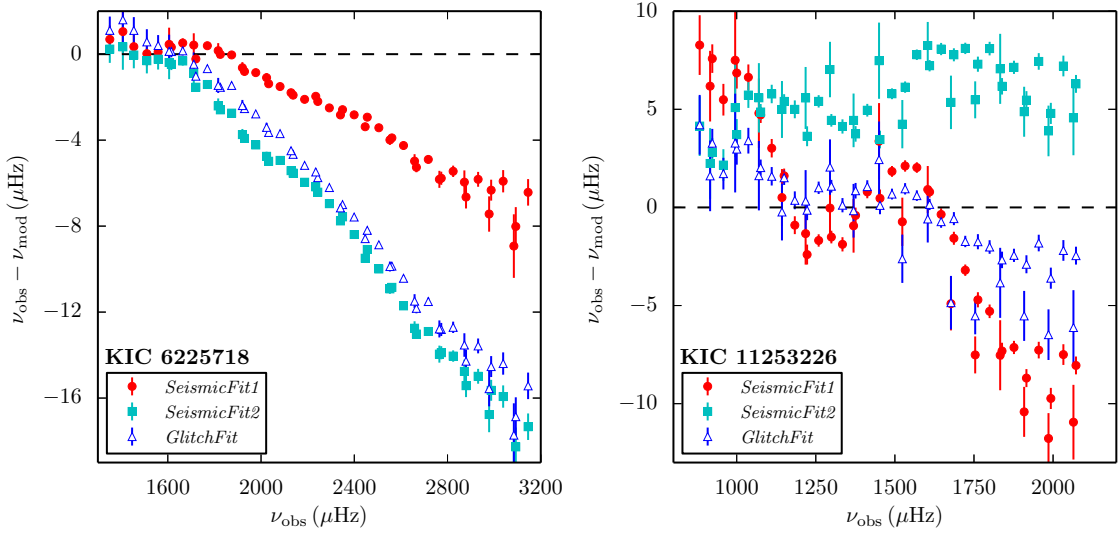}
\caption{Differences between the observed and a set of best-fit uncorrected model frequencies for KIC 6225718 and 11253226. 
The three different types of points in each panel correspond to three different sets of best-fit model frequencies obtained using
methods {\it SeismicFit1}, {\it SeismicFit2}, and {\it GlitchFit}.\label{fig9}}
\end{figure*}

The best-fit model obtained using only the spectroscopic and seismic data may not 
accurately reproduce the structure of the star, particularly the helium ionization layers, for the aforementioned reasons. In 
some cases, the signature of the mismatch of structure of the helium ionization layers in a star and the best-fit model can be 
seen directly in the difference between the observed and model frequencies. Figure~\ref{fig9} shows the difference between 
the observed and model frequencies for two such stars. The modulation on top of the surface term for the best-fit 
models obtained using {\it SeismicFit1} and {\it SeismicFit2} is due to the mismatch of the helium signature in the observed 
and model frequencies. The {\it GlitchFit} approach fits the glitch parameters and ensures that the observed and best-fit
model frequencies have similar helium signature, and consequently have either no modulation or smaller amplitude modulation
on top of the surface term. In this section, we illustrate using few individual stars how the glitch analysis helps us 
constrain the stellar structure better.

\subsection{Sun-as-a-star}
\begin{figure*}
\plotone{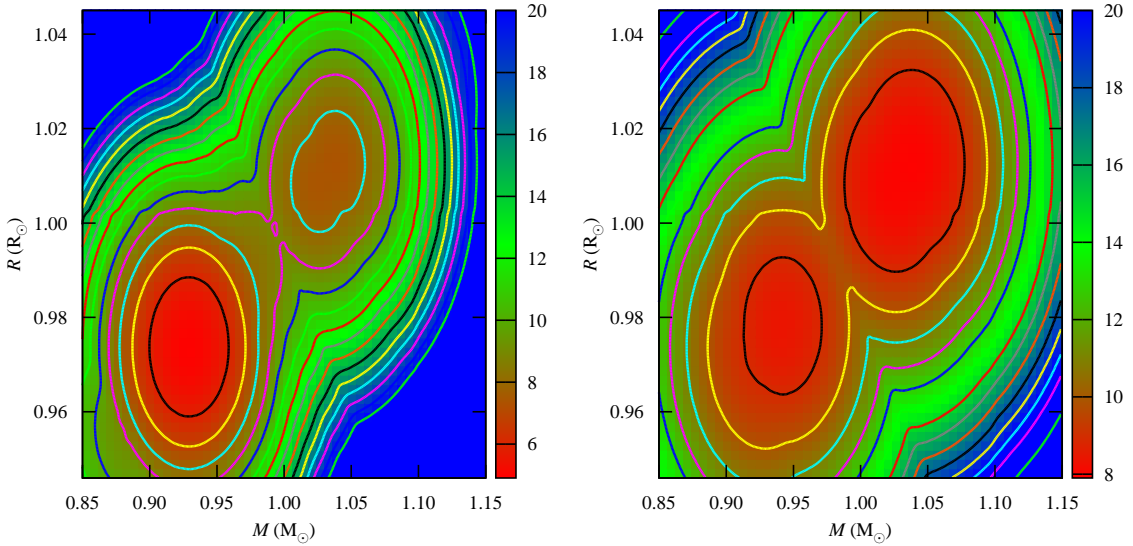}
\caption{Chi-square map as a function of mass and radius obtained from the ensemble of models constructed to fit the 
Sun-as-a-star data. The chi-square in the left panel includes only the spectroscopic and seismic observables ({\it SeismicFit1}),
while in the right panel it includes additional parameters associated with the helium signature ({\it GlitchFit}). The contours 
are uniformly spaced with $\Delta\chi^2 = 1$.\label{fig10}}
\end{figure*}

\citet{lund16} have also prepared data for the Sun with a noise level similar to the LEGACY sample to assess the results of 
their peak-bagging, and also to test the results of the modeling done by \citet{silv16}. We modeled the Sun in the same way as 
stars in the LEGACY sample with and without using the information from the glitch analysis. The first 
row of Table~\ref{tab2} lists the results obtained without using the information from glitch analysis. The mass and radius were 
found to be underestimated by about $2\sigma$. The surface helium abundance and the radial distance to the base of the convection 
zone, as obtained from the best-fit model, were also not consistent with the helioseismic determinations. The chi-square map 
obtained from the model ensemble suggested the possibility of a secondary solution with mass and radius closer to the solar 
value, as seen in the left panel of Figure~\ref{fig10}. 

\begin{figure*}
\plotone{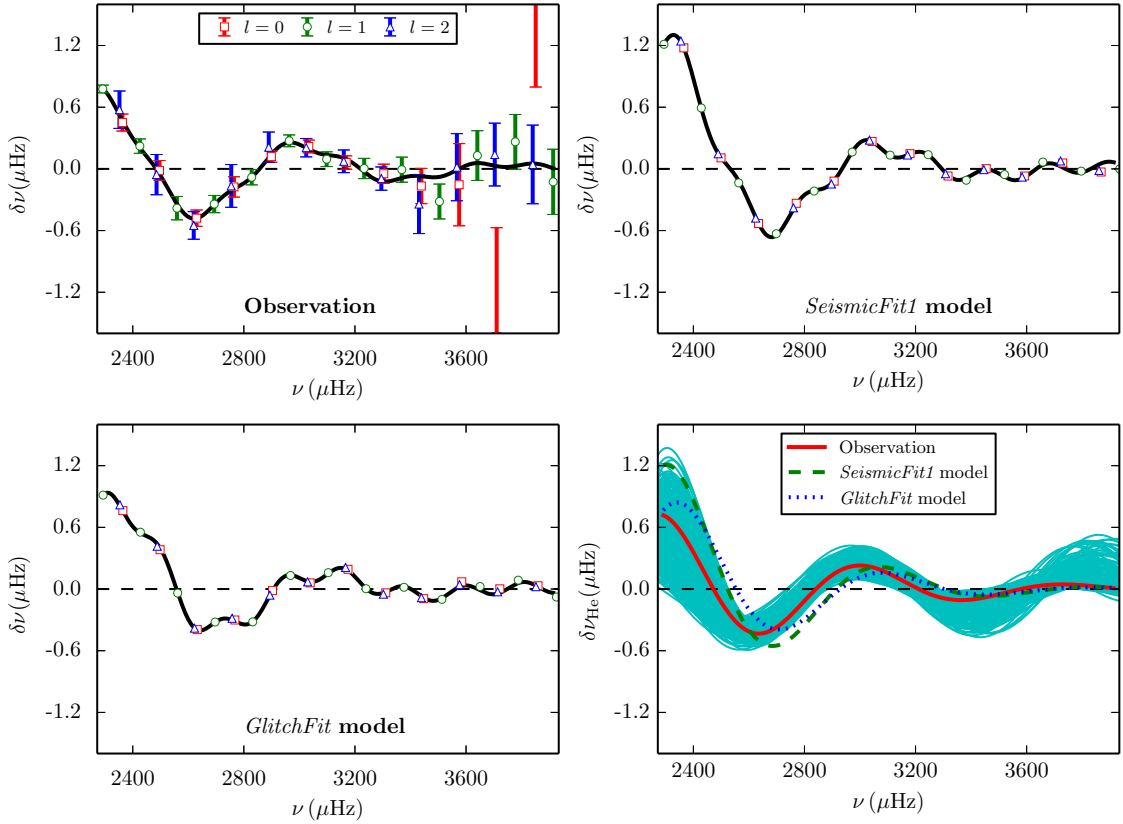}
\caption{Fit to the observed and best-fit model frequencies of Sun-as-a-star. The top left panel shows the fit to the observed 
frequencies, while the top right and bottom left panels show the fits to the best-fit model frequencies. The bottom right panel 
shows the corresponding helium glitch signatures. The helium signatures in the cyan color correspond to the fits to 500 
realizations of the observed frequencies.\label{fig11}}
\end{figure*}

A closer inspection of the results for the mass and surface helium abundance indicates that the problem could be due to the 
anti-correlation between the mass and initial helium abundance. A better constraint on the initial helium abundance can help in
such situations. The second row of Table~\ref{tab2} lists the results obtained using the supplementary information coming from
the glitch analysis. The mass and radius are both now in agreement with the solar mass and radius, and the values for the $Y$ 
and $R_{\rm CZ}$ are also closer to the helioseismic determinations. The right panel of Figure~\ref{fig10} shows the 
corresponding chi-square map. Note that the role of the primary and secondary solution has reversed. This is because the primary
solution in the left panel corresponds to a model that has significantly larger surface helium abundance than the solar helium 
abundance, and hence the corresponding oscillation frequencies have larger average amplitude of the helium signature, and
contributes significantly to the chi-square if the average amplitude is included in its definition. The secondary solution in 
the left panel, on the other hand, corresponds to a model that has surface helium abundance similar to the Sun, and hence their 
oscillation frequencies have similar average amplitude of helium signature, and contributes negligibly to the chi-square if the 
average amplitude is included in its definition. Figure~\ref{fig11} shows the fit to the observed as well as best-fit model 
frequencies, and compares their helium signatures. Apart from a small phase shift, the helium signature in the best-fit model 
frequencies obtained using {\it GlitchFit} reproduces the observed signature better than the best-fit model frequencies obtained 
using {\it SeismicFit1}. Recall that the phase of the helium signature was not included in the definition of 
$\chi^2_{\rm glitch}$, hence the possibility of a phase difference between the observed and model helium signature is not ruled 
out. This example clearly demonstrates that how the anti-correlation between the mass and initial helium abundance can lead to 
problems, and how the glitch analysis can help sort them out.

\begin{deluxetable*}{ccccrccccc}
\tabletypesize{\scriptsize}
\tablecaption{Physical parameters obtained with and without glitch parameters in the stellar model fitting.\label{tab2}}
\tablewidth{0pt}
\tablehead{
\colhead{Star} & \colhead{Method} & \colhead{$M$ (M$_\odot$)} & \colhead{$R$ (R$_\odot$)} & \colhead{$t$ (Gyr)} & 
\colhead{$L$ (L$_\odot$)} & \colhead{$\langle\rho\rangle$ (g cm$^{-3}$)} & \colhead{$Y$} & \colhead{$Y_i$} &
\colhead{$R_{\rm CZ}$ (R$_\odot$)}
}
\startdata
Sun-as-a-star & {\it SeismicFit1} & $0.93\pm0.03$ & $0.973\pm0.015$ &  $4.62\pm0.20$ & $0.937$ & $1.418\pm0.006$ & $0.277$ & $0.311$ & $0.725$\\
Sun-as-a-star &   {\it GlitchFit} & $1.03\pm0.04$ & $1.013\pm0.020$ &  $4.56\pm0.30$ & $0.991$ & $1.406\pm0.007$ & $0.237$ & $0.261$ & $0.717$\\
 KIC 8760414 & {\it SeismicFit1} & $0.83\pm0.02$ & $1.029\pm0.010$ & $13.82\pm0.40$ & $1.143$ & $1.070\pm0.003$ & $0.138$ & $0.203$ & $0.773$\\
 KIC 8760414 &   {\it GlitchFit} & $0.80\pm0.02$ & $1.018\pm0.010$ & $12.38\pm0.40$ & $1.136$ & $1.066\pm0.003$ & $0.183$ & $0.254$ & $0.750$\\
 KIC 6106415 & {\it SeismicFit1} & $1.05\pm0.04$ & $1.209\pm0.020$ &  $4.93\pm0.30$ & $1.699$ & $0.840\pm0.005$ & $0.221$ & $0.271$ & $0.937$\\
 KIC 6106415 &   {\it GlitchFit} & $1.05\pm0.04$ & $1.209\pm0.020$ &  $4.93\pm0.30$ & $1.699$ & $0.840\pm0.005$ & $0.221$ & $0.271$ & $0.937$
\enddata
\tablecomments{The solar helium abundance as obtained using helioseismology is, $Y = 0.248\pm0.003$ \citep{basu98}, while the 
radial distance of the base of the solar convection zone is, $R_{\rm CZ} = 0.713\pm0.001$ R$_\odot$ \citep{jcd91a}.}
\end{deluxetable*}

\subsection{KIC 8760414}
This star is one of the oldest and lowest metallicity star in the LEGACY sample ($[{\rm Fe}/{\rm H}] = -0.92\pm0.10$). It has 
been studied previously using {\it Kepler} data. For instance, \citet{math12} estimated the mass, radius, age, and the initial 
helium abundance of the star to be $0.81\pm0.01$ M$_\odot$, $1.02\pm0.01$ R$_\odot$, $13.35\pm0.38$ Gyr, and $0.220\pm0.018$, 
respectively, while \citet{metc14} found them to be $0.78\pm0.01$ M$_\odot$, $1.010\pm0.004$ R$_\odot$, $13.69\pm0.74$ Gyr, and 
$0.238\pm0.006$. We derived the physical properties of this star without using the information from the glitch analysis, and the 
results are listed in the third row of Table~\ref{tab2}. The small surface helium abundance is a result of both the small initial
helium abundance and large helium diffusion. In all the above determinations, the interesting point to note is that the initial 
helium abundance was found to be significantly sub-primordial \citep[$Y_{\rm P} = 0.2482 \pm 0.0007$;][]{stei10}, and the age to 
be close to the age of the universe \citep[$t_{\rm U} = 13.799 \pm 0.021$ Gyr;][]{plan16}.

\begin{figure}
\plotone{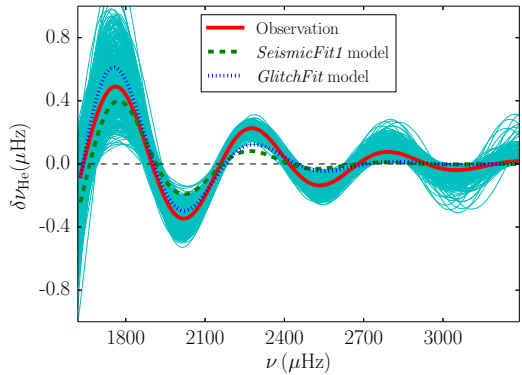}
\caption{Helium glitch signature in the observed and best-fit model frequencies of KIC 8760414. The helium signatures in the 
cyan color correspond to the fits to 500 realizations of the observed frequencies.\label{fig12}}
\end{figure}

The fourth row of Table~\ref{tab2} lists the parameters obtained using {\it GlitchFit}. The initial helium abundance is now 
greater than the amount of helium produced during the Big Bang nucleosynthesis, and is in line with the expectation from the 
helium-to-metal enrichment relation. It is interesting to note that the age of the star has come 
down significantly. Figure~\ref{fig12} compares the observed helium glitch signature with the corresponding signatures in the  
best-fit model frequencies. As we may expect from the too small surface helium abundance in the best-fit model obtained using 
{\it SeismicFit1}, the amplitude of the helium signature is smaller than the corresponding observed amplitude. The amplitude of 
the helium signature in the model frequencies obtained using {\it GlitchFit} is in better agreement with the observed one, 
particularly the amplitude averaged over the frequency range (better tracer of the helium abundance) are in much better agreement.

\subsection{KIC 6106415}
There are several stars in the LEGACY sample for which the results do not change on including the glitch parameters in the 
stellar model fitting. KIC 6106415 is an example of such a star. The results for this star are listed in the fifth and sixth rows 
of Table~\ref{tab2}, and are consistent with the results of earlier works \citep[see,][]{silv13}. The numbers in the two rows are
exactly the same because the best-fit models using two approaches turned out to be the same. Note from Table~\ref{tab2} 
that the masses, radii, and the ages were off only by about $2\sigma$ in the cases where secondary minimum was picked up by 
the {\it SeismicFit1} approach. This, in a way, justifies the other methods, which use only the spectroscopic and seismic 
data in the stellar model fitting. 

\section{Conclusions}
\label{sec:conc}
We fitted the signatures of the acoustic glitches in the oscillation frequencies of 66 main-sequence stars observed by 
{\it Kepler} satellite using two different methods, and derived the acoustic depths of the base of the convection zone and helium
ionization zone. We found that the signature from the He glitch is strong and the corresponding fit is robust for all stars, 
while the fit to the signature from the base of the convection zone is generally robust for the solar and sub-solar mass stars, 
but it is difficult to reliably fit its signature for super-solar mass stars. We fitted two different sets of best-fit 
model frequencies for all stars, and confirmed the findings of \citet{broo14} and \citet{verm14b} for the models of real 
stars that the fitted acoustic depth of the helium ionization zone correspond to the peak in the first adiabatic index between 
the first and second helium ionization zones.

We used the parameters associated with the helium glitch (average amplitude, width, and acoustic depth) together with the 
spectroscopic and seismic observables in the stellar model fitting to determine the stellar properties. The inclusion of the He 
glitch parameters puts tighter constraints on the stellar models, particularly on the initial helium abundance, and leads to a 
relatively more accurate set of stellar properties. This was demonstrated explicitly for the Sun-as-a-star and KIC 8760414 by
modeling them with and without the information from the glitch analysis. There are other stars in the sample with the bimodal 
distribution of chi-square (similar to what is shown in Figure~\ref{fig10}), for which the information from the glitch analysis 
helps constrain their properties better.

We studied the dependence of the various glitch parameters on the stellar parameters in the spirit of ensemble asteroseismology.
We found that the acoustic depths of the base of the convection zone and helium ionization zone are positively correlated. Since
the determination of $\tau_{\rm He}$ is reliable, we propose that the correlation be used as a guide to pick up the correct peak 
in the distribution of $\tau_{\rm CZ}$ (see, Figure~\ref{fig2} and \ref{fig3}), in cases when it has multiple peaks. The average
amplitude of the helium signature increases not only with the helium abundance but also with the mass (hence the effective 
temperature), therefore a careful calibration is required to estimate the envelope helium abundance. The helium abundance obtained
using the information from the glitch analysis is expected to be more reliable, and a detailed analysis using different
methods, including its determination from the calibration of the observed average amplitude of the helium signature, will be 
presented in future. 

\acknowledgments
Funding for the Stellar Astrophysics Centre is provided by The Danish National Research Foundation (Grant DNRF106). The research 
was supported by the ASTERISK project (ASTERoseismic Investigations with SONG and Kepler) funded by the European Research Council
(Grant agreement no.: 267864). Authors thank J. Christensen-Dalsgaard and G. Houdek for a careful reading of the manuscript. KV 
thanks S.~M. Chitre for his support at CEBS, where the first draft of the paper was written. KR and AM acknowledge support from 
the NIUS program of HBCSE (TIFR). SB is partially supported by NSF grant AST-1514676 and NASA grant NNX16A109G. MNL acknowledges 
the support of The Danish Council for Independent Research | Natural Science (Grant DFF-4181-00415). VSA acknowledges support 
from VILLUM FONDEN (research grant 10118).


\end{document}